\documentclass[%
 reprint,
superscriptaddress,
nofootinbib,
 amsmath,amssymb,
 aps,
pra,
]{revtex4-2}

\usepackage[utf8]{inputenc}
\usepackage[english]{babel}

\usepackage{amsmath}
\usepackage{amssymb}
\usepackage{bbm}
\usepackage{bm}
\usepackage{braket}
\usepackage{graphicx}
\usepackage{dsfont}

\usepackage[usenames,dvipsnames]{xcolor}
\usepackage[colorlinks=true,citecolor=MidnightBlue,linkcolor=MidnightBlue,urlcolor=MidnightBlue]{hyperref}
 
\begin{document}

\title{Generating Greenberger-Horne-Zeilinger States Using Multiport Splitters}

\author{Daniel Bhatti}
\affiliation{Institute for Functional Matter and Quantum Technologies, University of Stuttgart, 70569 Stuttgart, Germany}
\affiliation{Center for Integrated Quantum Science and Technology (IQST), University of Stuttgart, 70569 Stuttgart, Germany}
\author{Stefanie Barz}
\affiliation{Institute for Functional Matter and Quantum Technologies, University of Stuttgart, 70569 Stuttgart, Germany}
\affiliation{Center for Integrated Quantum Science and Technology (IQST), University of Stuttgart, 70569 Stuttgart, Germany}

\begin{abstract}
Symmetric multiport splitters are versatile tools in optical quantum information processing.
They can be used for studying multiparticle scattering, studying distinguishability and mixedness, and also for the generation of multipartite entangled quantum states.
Here, we show that $N$-photon $N$-mode Greenberger-Horne-Zeilinger (GHZ) states can be generated using symmetric multiport beam splitters. 
Varying the input states' internal degrees of freedom and post-selecting onto certain photon-number distributions allows the probabilistic generation of GHZ states with arbitrary photon numbers.
We present two novel schemes, one for odd and one for even numbers of photons, to generate GHZ states using symmetric multiport splitters and compare them to a strategy utilizing a $2N$-port network as well as the standard post-selection method.
\end{abstract}

\maketitle

\section{Introduction}

Multiparty Greenberger-Horne-Zeilinger (GHZ) states lie at the heart of many theoretical and experimental investigations~\cite{Hillery1999,Epping2017,Grasselli2018,Wang2018,Omran2019,Murta2020,Hahn2020,Thalacker2021,Pogorelov2021,Mooney2021,Thomas2022,Meyer-Scott2022}.
It is their characteristic properties, the maximal $N$-qubit entanglement and the maximal correlation between all qubits, which not only lead to a sharp contradiction between the classical theory of local hidden variables and quantum mechanics \cite{Greenberger1990,Pan2000}, but also make them ideal for quantum communication~\cite{Murta2020}.
For example, they are an important resource for (anonymous) conference key agreement~\cite{Epping2017,Grasselli2018,Hahn2020,Thalacker2021}, and secret sharing \cite{Hillery1999}, but can also be used as building blocks to fuse them into larger cluster states~\cite{Omkar2021}. Recently, higher-dimensional, hybrid- and hyper-entangled GHZ states have been investigated~\cite{Erhard2020,Zhang2020,Xia2012}, enabling superdense coding, hybrid quantum networks, and improved identification of entangled states, respectively.

Thus, generating GHZ states is highly relevant and it is particularly important to find easily accessible methods and protocols.
Experimentally, the generation of GHZ states has been demonstrated within a versatile range of platforms:
in photonic systems $14$-qubit GHZ states using 14 photons~\cite{Thomas2022}, and 18-qubit GHZ state using 6 photons~\cite{Wang2018} have been shown.
In the case of Rydberg atoms, ions, and superconducting qubits, 20-qubit, 24-qubit, and 27-qubit GHZ states have been demonstrate, respectively~\cite{Omran2019,Pogorelov2021,Mooney2021}.
In particular, for networked quantum applications such as communication protocols the generation of large photonic GHZ states is important~\cite{Murta2020}.

One way of generating highly entangled photonic multiparty quantum states
is to post-select for $N$-photon coincidences in $N$ spatial modes.
For example, one can fuse polarization-entangled photon pairs into polarization-encoded GHZ states using polarizing beam splitters and $N$-photon post-selection only~\cite{Thalacker2021}. These kinds of setups reach relatively high success probabilities, but can generate a specific state only.

From far-field interference setups it is known that all symmetric states can be generated~\cite{Maser2010},
including GHZ states~\cite{Bastin2009,Hossein-Nejad2009}. Such setups benefit from the fact that the generated entangled state, even the entanglement class, can be changed by simply adjusting, e.g., the input photons' internal degrees of freedom, however, with very low success probabilities~\cite{Hossein-Nejad2009}.

\begin{figure}[b]
	\centering
		\includegraphics[width=0.7\columnwidth]{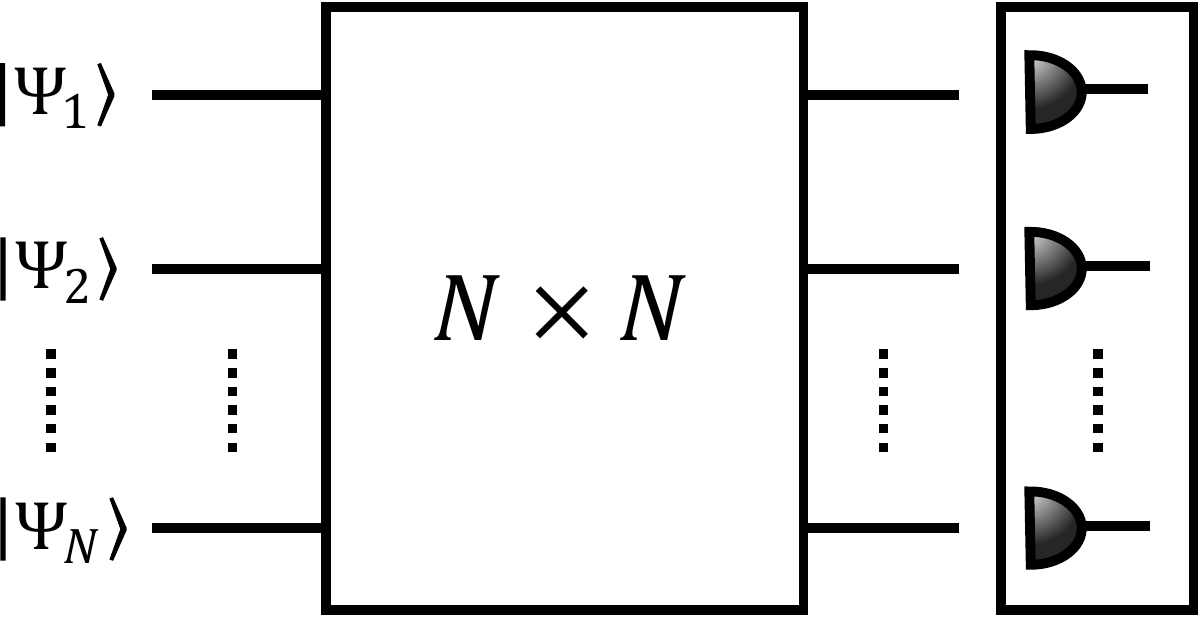}
	\caption{Unitary multiport splitter $U_{N}$ with $N$ input and $N$ output ports. Usually, each input port $k$ contains a single photon in the state $\ket{\Psi_{k}}$. Entangled multiparty quantum states can then be generated from post-selecting $N$-photon coincidences in the $N$ output modes with the help of single-photon detectors.}
	\label{fig:NPort_basic}
\end{figure}

A flexible possibility to obtain higher probabilities is to use multiport splitters (MSs) (see Fig.~\ref{fig:NPort_basic})~\cite{Pryde2003,Lim2005,Maser2010,Kiesel2010,Kasture2018,Paesani2021,Lee2022,Bell2022}.
There, single photons are sent into a MS and specific output configurations are considered, e.g., the ones containing exactly one photon per output port.
In contrast to networks specifically designed to generate a particular entangled state \cite{Ju2019,Blasiak2019,Kim2020,Lee2022,Blasiak2021}, MSs -- like far-field interference setups -- allow for the generation of different entangled states even belonging to different entanglement classes by simply adjusting the input photons' internal degrees of freedom~\cite{Lim2005,Kumar2022}.
While the role of the internal degrees of freedom in multi-photon interferences and its connection to the general counting statistics and entanglement production has been investigated~\cite{Lim2005,Menssen2017,Dittel2018,Jones2020,Minke2021}, it still remains an open question how to find appropriate input states for the generation of specific (entangled) output states.

In this paper, we study the generation of $N$-photon $N$-mode GHZ states in symmetric MSs (SMSs) and prove that the generation of GHZ states is possible for arbitrary $N$ using post-selection on photon numbers.
We present the input states for arbitrary odd and even $N$, and show that the post-selected output states with a single photon in each of the spatial output modes can only be of the form of GHZ states.
For small numbers of photons and input ports, meaning up to four, this behavior has already been studied~\cite{Shih1988,Lim2005,Kumar2022}.
However, in contrast to W states for which a general scheme already exists~\cite{Lim2005}, a general scheme for the generation of GHZ states using SMSs has still been missing. Our results close this gap and, thereby, open up the possibility to employ any SMS to distribute entangled states of distinct entanglement classes, which can directly be used in an arbitrary $N$-party quantum network.

The paper is structured as follows: in Sec.~\ref{sec:Theory}, we start by discussing the generalized Hong-Ou-Mandel effect using SMSs~\cite{Lim2005IOP} to introduce the general system, the post-selection mechanism, and the mathematical tools. In particular, we introduce the zero transmission law~\cite{Tichy2010}, which helps us to identify suppressed photon output distributions. In Sec.~\ref{sec:2NNetwork} we discuss a possible scheme to generate $N$-photon GHZ states using a MS with $2N$ inputs and outputs, respectively.
In Secs.~\ref{sec:MultiModeGHZ} and \ref{sec:MultiModeGHZEven} we present our solutions to generate $N$-photon GHZ states from symmetric $N$-port splitters for odd $N$ and even $N$, and explain why the chosen input states can only lead to photon distributions belonging to a GHZ state. We discuss the success probabilities and the scaling of the different schemes in Sec.~\ref{sec:Discussion} and conclude in Sec.~\ref{sec:Conclusion}.

\section{Symmetric Multiport Splitters}
\label{sec:Theory}

In this work, we discuss the behavior of $N$ photons impinging on unitary MSs to generate entangled states. For this, we focus on qubits and the generation of $N$ qubit GHZ states.
Note that, in principal, the discussion could easily be extended to qudits to investigate the generation of entangled $N$-qudit states from MSs.

\begin{figure}[t]
	\centering
		\includegraphics[width=0.8\columnwidth]{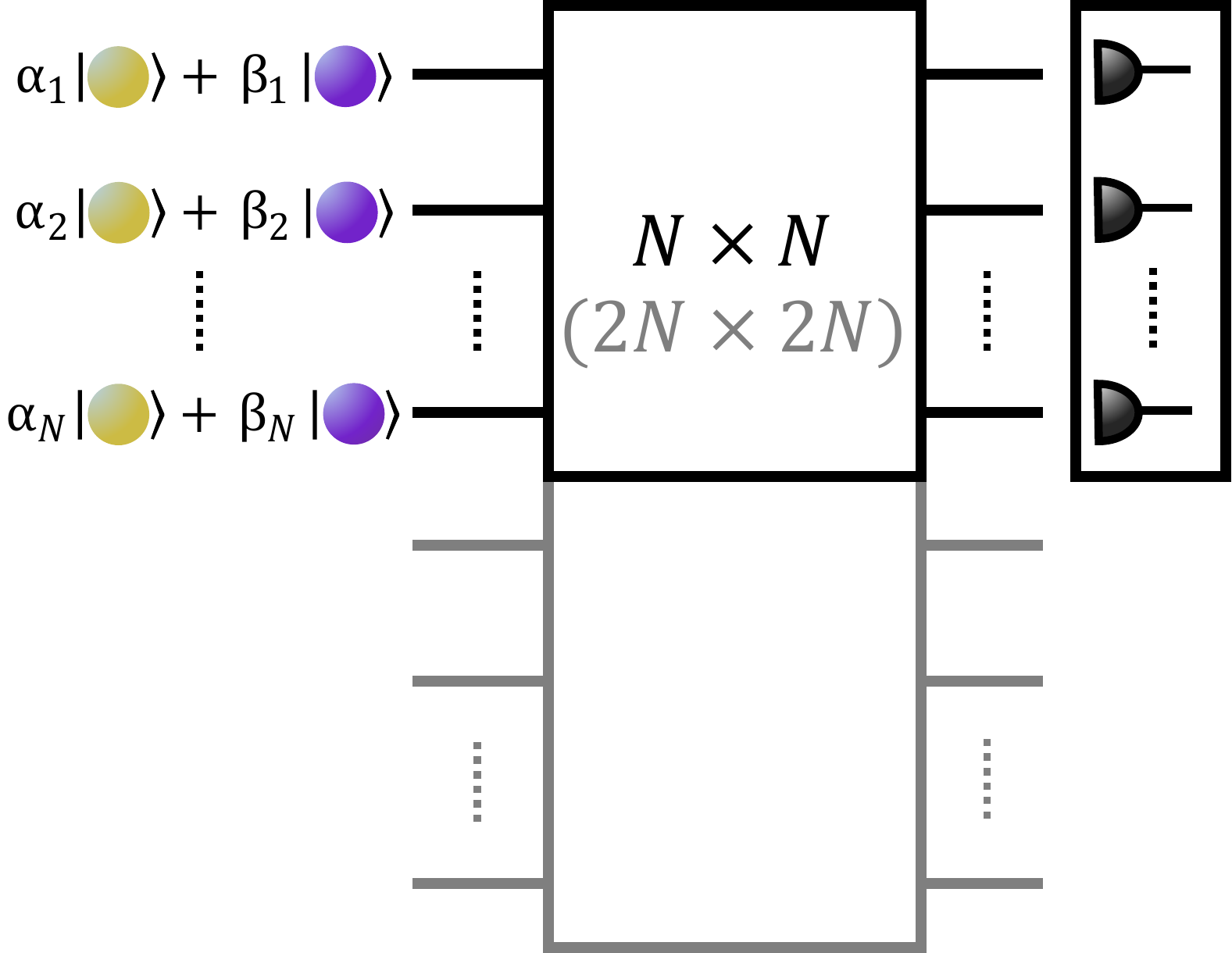}
	\caption{Unitary networks considered in Secs.~\ref{sec:2NNetwork} ($2N\times 2N$) and \ref{sec:MultiModeGHZ} ($N\times N$) to generate GHZ states. $N$ photonic qubits prepared in a superposition state each are being sent into the first $N$ input modes of the $N$-port or $2N$-port network, respectively. Post-selection is then used to detect $N$-mode $N$-photon coincidences in the first $N$ output modes of the networks. For the $2N$-port network, the second $N$ input ports and the second $N$ output ports remain unused, i.e., empty.}
	\label{fig:SetupNetwork}
\end{figure}

The photonic $N$-qubit input state of a unitary MS with $N$ (spatial) input modes and $N$ (spatial) output modes (see Fig.~\ref{fig:SetupNetwork}), can be written in the form \cite{Lim2005IOP}
\begin{align} \label{eq:PsiIn}
	\ket{\Psi_{\text{in}}} & = \prod_{k=1}^{N} \left( \alpha_{\mu,k} a_{\mu,k}^{\dagger}+\alpha_{\eta,k} a_{\eta,k}^{\dagger} \right) \ket{0} \nonumber \\
	& = \prod_{k=1}^{N} \left( \alpha_{\mu,k} \ket{\mu}_{k}+\alpha_{\eta,k} \ket{\eta}_{k} \right) ,
\end{align}
where $\alpha_{\mu,k}$ and $\alpha_{\eta,k}$ are complex coefficients with $\left| \alpha_{\mu,k} \right|^{2} + \left| \alpha_{\eta,k} \right|^{2} = 1$ for all $k=1,\ldots,N$. Here, the creation (annihilation) of a photonic qubit in the $k$th-mode with the internal degree of freedom $F=\mu,\eta$, is described by the operator $a_{F,k}^{\dagger}$ ($a_{F,k}$), with the respective single-photon quantum state $\ket{F}_{k}$, and ${\vphantom{\braket{\mu|\eta}}}_{k}\!\!\braket{\mu|\eta}_{k}=0$. 
In general, one can utilize any internal degree of freedom that can realize a photonic qubit (see Appendix~\ref{sec:InternalDoF}), e.g., polarization or time-bin encoding, and we can define $\ket{\pm}_{k}=[\ket{\mu}_{k}\pm \ket{\eta}_{k}]/\sqrt{2}$, and $\ket{R/L}_{k}=[\ket{\mu}_{k}\pm i \ket{\eta}_{k}]/\sqrt{2}$.
Note that in the following, we only specify the internal degree of freedom for distinguishable photons whereas we do not explicitly state the internal degrees of freedom for indistinguishable photons.

In general, one can use a unitary matrix $U$ to describe the transformation of any $N$-mode input state into its respective $N$-mode output state. The matrix elements $U_{kl}$ ($k,l=1,2,\ldots,N$), thereby, denote the transition for an input photon going from input mode $k$ to output mode $l$. The mathematical relation between input creation operators $a_{F,k}^{\dagger}$ and output creation operators $b_{F,l}^{\dagger}$ is given by
\begin{align}
	b_{F,l}^{\dagger} = \sum_{k=1}^{N} U_{lk} a_{F,k}^{\dagger}  ,
\end{align}
which leads to the following transition relation mapping input to output operators:
\begin{align} \label{eq:TransformationRelation}
	a_{F,k}^{\dagger} \rightarrow \sum_{l=1}^{N} U_{kl} b_{F,l}^{\dagger}  .
\end{align}

In the following, we focus on SMSs, which have a uniform output distribution. The respective unitary transformation is identical to a discrete Fourier transform (DFT) \cite{Lim2005IOP}
\begin{align}
	U_{kl} = \frac{1}{\sqrt{N}} \omega_{N}^{(k-1)(l-1)} ,
\label{eq:Fourier}
\end{align}
with $\omega_{N}=\exp\!\left(\text{i} 2\pi/N\right)$ being the $N$th root of unity.

\begin{figure*}[t]
	\centering
		\includegraphics[width=1.9\columnwidth]{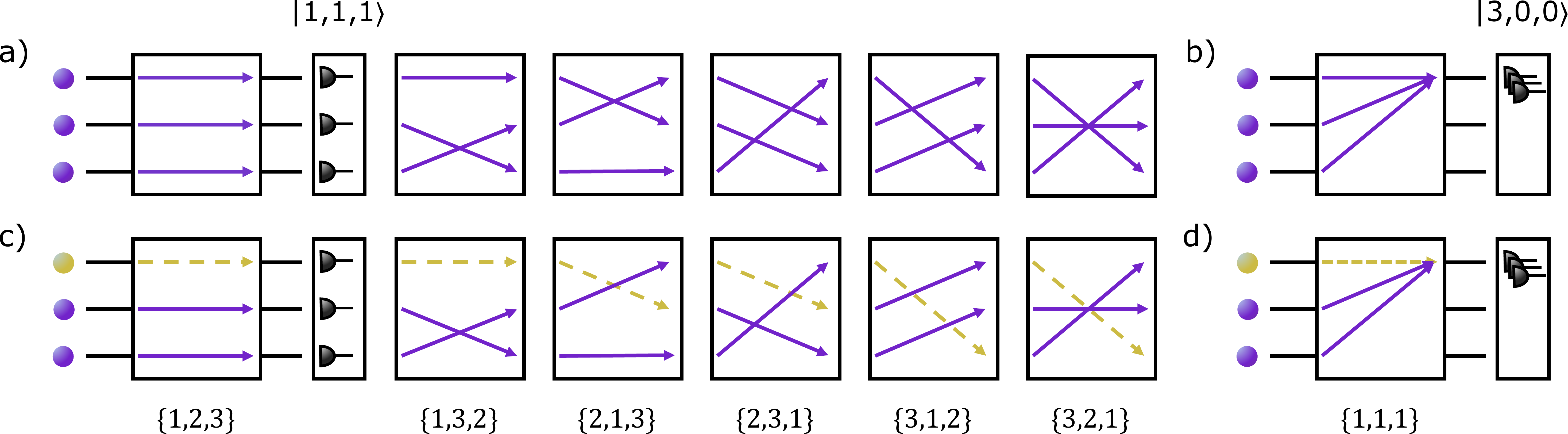}
	\caption{All possibilities how three input photons can be detected at the output of a symmetric three-port network and contribute to different post-selection schemes. Post-selecting for one photon in each of the output modes, i.e., the photon-number state $\ket{1,1,1}$, six different possibilities exist for a) three indistinguishable photons and c) two indistinguishable photons and one distinguishable photon. In both cases different possibilities concur with the permutations of $\sigma=\{1,2,3\}$ as described in the main text. However, c) contains three distinguishable configurations and post-selection generates an entangled output state. Post-selecting all three photons in the same output mode, i.e., the state $\ket{3,0,0}$, only a single possibility, i.e., $\{1,1,1\}$, exists for b) three indistinguishable photons and d) two indistinguishable photons and one distinguishable photon.}
	\label{fig:QuantumPaths}
\end{figure*}

We start with an $N$-photon input state for perfectly indistinguishable photons and one photon per input mode [see Eq.~(\ref{eq:PsiIn}) with all $\alpha_{\eta,k}=0$], i.e.,
\begin{align}
	\ket{\tilde{\Psi}_{\text{in}}}  = \prod_{k=1}^{N} a_{k}^{\dagger}\ket{0} \equiv \ket{1,1,\ldots ,1} ,
\label{eq:Psi_in}
\end{align}
where $\ket{n_{1},n_{2},\ldots,n_{N}}=\ket{n_{1}}_{1}\otimes\ket{n_{2}}_{2}\otimes\cdots\otimes\ket{n_{N}}_{N}$ denotes the photon-number state in the Fock basis. We now explain the idea of post-selection.

By making use of the transformation relation induced by the DFT [see Eqs.~(\ref{eq:TransformationRelation}) and (\ref{eq:Fourier})], i.e.,
\begin{align}
	a_{k}^{\dagger} \rightarrow \sum_{l=1}^{N} U_{kl} b_{l}^{\dagger} =   \frac{1}{\sqrt{N}} \sum_{l=1}^{N} \omega_{N}^{(k-1)(l-1)} b_{l}^{\dagger} ,
\label{eq:TransformationRelationDFT}
\end{align}
the input state $\ket{\tilde{\Psi}_{\text{in}}}$ given by Eq.~(\ref{eq:Psi_in}) is transformed into the complete output state $\ket{\tilde{\Psi}_{\text{out}}}$, i.e.,
\begin{align}
	\ket{\tilde{\Psi}_{\text{in}}} \rightarrow \ket{\tilde{\Psi}_{\text{out}}} 
	& = \prod_{k=1}^{N} \sum_{l=1}^{N} U_{kl} b_{l}^{\dagger} \ket{0} \nonumber \\
	 & = \frac{1}{\sqrt{N}^{N}} \prod_{k=1}^{N} \sum_{l=1}^{N} \omega_{N}^{(k-1)(l-1)} b_{l}^{\dagger} \ket{0} .
\label{eq:Psi_out}
\end{align}

Since we have a single photon per input mode $k$ which each can go to any of the output modes $l$, Eq.~(\ref{eq:Psi_out}) contains the product over all input modes $k$ and the sum over all output modes $l$. In combination, this coincides with the sum over all possibilities for how the photons can transit from input modes to output modes and can be rearranged into two steps. 

In the first step, we investigate the photon-number output states $\ket{n_{1},n_{2},\ldots,n_{N}}$, with $0\leq n_{l}\leq N$ ($l=1,2,\ldots,N$), and $\sum n_{l}=N$. Without post-selection one needs to consider all possible photon-number states to find the complete output state given by Eq.~(\ref{eq:Psi_out}). However, post-selecting for a single photon-number output state -- usually $\ket{1,1,\ldots,1}$ -- one can concentrate on this specific state and calculate its success probability in the next step.
Note that in the case of indistinguishable photons the different output states $\ket{n_{1},n_{2},\ldots,n_{N}}$ are solely determined by the number of photons per output mode.
In the case of partially distinguishable photons where the internal degrees of freedom of the photons are distinguishable, meaning the scalar products of pairs of photons $<1$ (see Appendix~\ref{sec:InternalDoF}), we further need to discuss the photons' internal states. This often leads to the generation of entanglement in post-selection schemes.

In the second step, we calculate the success probability of the post-selected output state $\ket{n_{1},n_{2},\ldots,n_{N}}$,
by summing over all possibilities how the $N$ input photons can be detected at the output modes of the SMS
(see Fig.~\ref{fig:QuantumPaths}).
From $\ket{n_{1},n_{2},\ldots,n_{N}}$ we already know, that $n_{1}$ photons go to output mode 1, $n_{2}$ photons go to output mode 2, \ldots, and $n_{N}$ photons go to output mode $N$.
This defines the totally ordered set of all output modes $\sigma=\{l_{1},l_{2},\ldots ,l_{N}\}$, with $1\leq l_{i}\leq l_{j} \leq N$ ($i< j$), which contains $n_{1}$ 1s, $n_{2}$ 2s, \ldots, and $n_{N}$ $N$s. All possible $N$-photon contributions belonging to this specific state can then be found from all possible permutations $\mathcal{P}_{\sigma}$ of $\sigma$ (see Fig.~\ref{fig:QuantumPaths}).
For example, in the case of exactly one photon per output mode, i.e., $\ket{1,1,\ldots,1}$, $\sigma=\{1,2,\ldots ,N\}$, which has $N!$ permutations; in the case that all photons exit through the same output mode $l$, i.e., $\ket{0,\ldots,0,N,0,\ldots,0}$, $\sigma=\{l,l,\ldots,l\}$, which has only $1$ permutation.

Finally, one can also write the post-selection process in the form of a projection, i.e.,
\begin{align}
	P=\ket{n_{1},n_{2},\ldots,n_{N}}\!\!\bra{n_{1},n_{2},\ldots,n_{N}} ,
\end{align}
which leads to [see Eq.~(\ref{eq:Psi_out})]
\begin{align}
	\ket{\tilde{\Psi}_{\text{out}}} & =  P \ket{\tilde{\Psi}_{\text{out}}} + (\mathds{1}-P)\ket{\tilde{\Psi}_{\text{out}}} \nonumber \\
	& \equiv \ket{\tilde{\psi}_{\text{out,ps}}} + \ket{\tilde{\psi}_{\text{out,rest}}} .
\end{align}
Here, $\ket{\tilde{\psi}_{\text{out,ps}}}$ denotes the post-selected quantum state, while $\ket{\tilde{\psi}_{\text{out,rest}}}$ denotes the rest, i.e., the non-post-selected quantum state. Both states are not normalized to unity, but to their respective probabilities to occur.

With this, we can now calculate the general post-selected state~\cite{Lim2005IOP}
\begin{align}
	\ket{\tilde{\psi}_{\text{out,ps}}} & = \sum_{\mathcal{P}_{\sigma}} \left[ \prod_{k=1}^{N} U_{k \sigma(k)} b_{\sigma(k)}^{\dagger} \right] \ket{0} \nonumber \\
	& = \frac{1}{\sqrt{N}^{N}} \sum_{\mathcal{P}_{\sigma}} \left[ \prod_{k=1}^{N} \omega_{N}^{(k-1)(\sigma(k)-1)} b_{\sigma(k)}^{\dagger} \right] \ket{0} ,
\label{eq:Psi_out_111}
\end{align}
where $\sum_{\mathcal{P}_{\sigma}}$ denotes the sum over all permutations of the output modes $\sigma=\{l_{1},l_{2},\ldots ,l_{N}\}$, and $\sigma(k)$ denotes the $k$th entry of the permutation. Together with the product over the different input modes $\prod_{k}$ this again corresponds to 
all possibilities for how the photons can transit from the input modes to the output modes and form the post-selected state.
Note that one can either permute the input modes and keeping the output modes fixed. Or, alternatively, we can permute the output modes and keep the input modes fixed.
We will make use of this possibility throughout the following investigations and define $\delta=\{k_{1},k_{2},\ldots ,k_{N}\}$ as the totally ordered set of input modes.

We now calculate the success probability $P_{\text{ps}}$ for the post-selected event to occur. This probability can be obtained from the post-selected state of Eq.~(\ref{eq:Psi_out_111}) and is given by
\begin{align}
\label{eq:P_ps}
	P_{\text{ps}} & = |\!| \ket{\tilde{\psi}_{\text{out,ps}}} |\!|^{2} \nonumber \\
	& = \frac{1}{N^{N}} \big| \sum_{\mathcal{P}_{\sigma}} \prod_{k=1}^{N} \omega_{N}^{(k-1)(\sigma(k)-1)} \big|^{2},
\end{align}
where $|\!| \ket{\Psi} |\!|^{2} = \braket{\Psi | \Psi}$ defines the norm of the state $\ket{\Psi}$.

It has been shown that post-selecting for a single photon per output mode, i.e., projecting onto the state $\ket{1,1,\ldots,1}$, a generalized Hong-Ou-Mandel (HOM) effect occurs and the probability given by Eq.~(\ref{eq:P_ps}) vanishes for even $N$~\cite{Lim2005IOP}.
Eq.~(\ref{eq:P_ps}) has also been analyzed for arbitrary output distributions and it has been shown that an input state given by Eq.~(\ref{eq:Psi_in}) can only produce output distributions, which fulfill the condition~\cite{Tichy2010}
\begin{align}
	\sum_{i=1}^{N} l_{i} = 0 \ \text{mod}(N) .
\label{eq:ZTL}
\end{align}
This result is called the zero transmission law (ZTL)~\cite{Tichy2010} and will be used in the following investigations.
Note that the same law has been derived 
for multi-photon interferences in free space~\cite{Classen2016,Bhatti2018}.

We can now move on to the discussion of the generation of $N$-mode GHZ states.

\section{N-Photon GHZ State Generation}

In this section we investigate three different MSs, i.e., a non-symmetric $2N$-port splitter, a symmetric $N$-port splitter for odd $N$, and a symmetric $N$-port splitter for even $N$.
We show that by choosing specific qubit-encoded input states and post-selecting for $N$-photon coincidences $N$-photon $N$-mode GHZ states can be generated for arbitrary $N$.

\subsection{2N-port Network}
\label{sec:2NNetwork}

Let us start by discussing the generation of $N$-mode $N$-photon GHZ states using $2N$-port networks of the form~\cite{Kasture2018}
\begin{align}
U_{2N} = 	\begin{pmatrix}
A & (\mathds{1}-AA^{\dagger})^{1/2} \\
(\mathds{1}-A^{\dagger}A)^{1/2} & -A^{\dagger} 
\end{pmatrix} ,
\label{eq:2NNetwork}
\end{align}
where $A$ denotes a nonunitary transformation, with all entries $A_{kl} = 1/N$ ($k,l=1,2,\ldots,N$).
By sending in $N$-photons into the first $N$ input modes and at the same time post-selecting for $N$-photon coincidences exclusively in the first $N$ output modes (see Fig.~\ref{fig:SetupNetwork}), one can reduce the effective transformation to $A$ and we find the following transformation relation for an input photon in the $k$th input mode:
\begin{align} \label{eq:TransformationRelationNonUnitary}
	a_{F,k}^{\dagger} \rightarrow \frac{1}{N} \sum_{l=1}^{N} b_{F,l}^{\dagger}  .
\end{align} 
This scheme has been suggested to generate symmetric $N$-mode Dicke states using MSs~\cite{Kasture2018} and should, in our case, yield the wanted GHZ states.

To this aim we make use of the $N$-photon state~\cite{Bastin2009,Hossein-Nejad2009}
\begin{align}
	&\ket{\Psi_{2N,\text{in}}} = \frac{1}{\sqrt{2}^{N}}\prod_{k=1}^{N} \left[ a_{\mu,k}^{\dagger} + e^{\text{i}\theta_{k}}a_{\eta,k}^{\dagger} \right] \ket{0} ,
\label{eq:InputState2}
\end{align}
with $\theta_{k} = (k-1) 2\pi/N$, $k=1,2,\ldots,N$.
In Refs.~\cite{Bastin2009,Hossein-Nejad2009} it has been shown that choosing this input state along with a symmetric detection strategy corresponding to Eq.~(\ref{eq:TransformationRelationNonUnitary}) $N$-qubit GHZ states can be generated using $N$ single-photon sources and $N$-photon post-selection.

\begin{figure}[b]
	\centering
		\includegraphics[width=0.7\columnwidth]{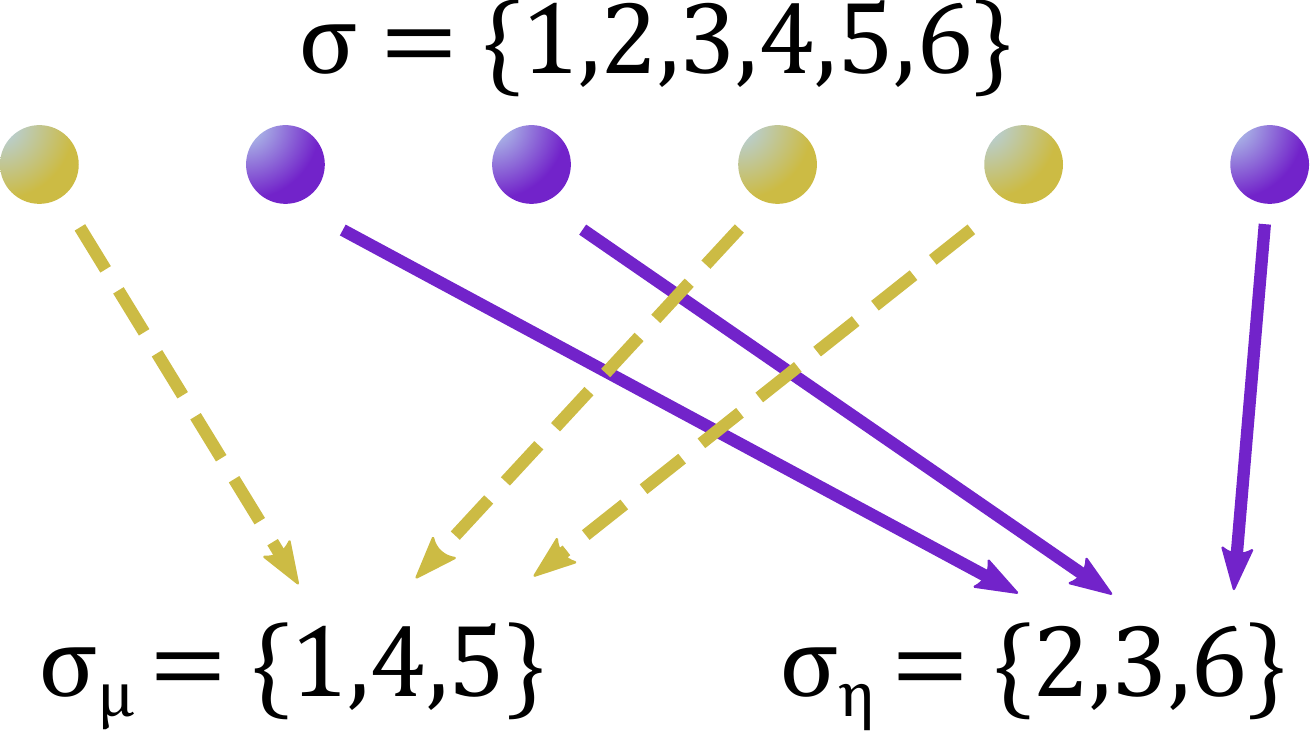}
	\caption{Example for how the complete set of output modes $\sigma = \{1,2,\ldots,N \}$ of $N$ photons is split up into two subsets $\sigma_{\mu}$ and $\sigma_{\eta}$ containing only the output modes of the $N_{1}$ photons in the state $\ket{\mu}$ and the $N_{2}$ photons in the state $\ket{\eta}$, respectively. In the figure, for $N=6$, we show the case, where $\sigma$ contains $N_{1}=N_{2}=3$ photons in $\ket{\mu}$ (yellow, dashed arrows) and $\ket{\eta}$ (purple, solid arrows), which leads to $\sigma_{\mu} = \{1,4,5\}$ and $\sigma_{\eta} = \{2,3,6\}$, with $\sigma=\sigma_{\mu} \cup \sigma_{\eta}$.}
	\label{fig:PhotonDistributions}
\end{figure}

For the $2N$-network we find the output state in the first $N$ output modes of the MS
\begin{align}
	& \ket{\Psi_{2N,\text{out}} } = \frac{1}{\sqrt{2}^{N}N^{N}} \prod_{k=1}^{N} \left[ \sum_{l=1}^{N} b_{\mu,l}^{\dagger} + e^{\text{i}\theta_{k}}\sum_{l=1}^{N} b_{\eta,l}^{\dagger} \right] \ket{0} ,
\end{align}
where we have inserted the normalized transformation relation defined by Eq.~(\ref{eq:TransformationRelationNonUnitary}) into Eq.~(\ref{eq:InputState2}). Now we can use the post-selection rule, i.e., we want to measure exactly a single photon per output mode and we thereby post-select for the photon-number state $\ket{1,1,\ldots,1}$. Since the photons can be in distinguishable internal states $\ket{\mu}$ and $\ket{\eta}$, we need to analyze every specific output distribution of $N_{1}$ photons in the state $\ket{\mu}$ and $N_{2}$ photons in the state $\ket{\eta}$, with $N=N_{1}+N_{2}$, to determine the exact form of the post-selected state. The interference term of each possible output distribution reads [see Eq.~(\ref{eq:P_ps})]
\begin{align}
	&	\sum_{\mathcal{P}_{\delta}} \left[\, \prod_{l=1}^{N_{1}}  \times \prod_{l'=1}^{N_{2}}  e^{\text{i}\theta_{\delta(N_{1}+l')}} \right]   .
\label{eq:PostselectHV3}
\end{align}
where we now sum over all possible permutations $\mathcal{P}_{\delta}$ of the set of input modes $\delta=\{1,2,\ldots,N\}$. Note that the product over the $N$ output modes is now divided into two products, one for the $N_{1}$ photons in the state $\ket{\mu}$ and one for the $N_{2}$ photons in the state $\ket{\eta}$. By applying the ZTL [see Eq.~(\ref{eq:ZTL})] one sees that this sum can only be unequal to zero if
\begin{align}
	N_{2} =  0 \ \text{mod}(N) .
\end{align}
This condition can only be fulfilled for $N_{2}= 0$ or $N_{2}=N$. The final post-selected state is then a GHZ state of the form
\begin{align}
	\ket{\psi_{2N,\text{ps}}} =& \frac{N!}{\sqrt{2}^{N}N^{N}} \left[ \ket{\mu}_{1}\ket{\mu}_{2}\ldots\ket{\mu}_{N} \right. \nonumber \\ 
	& \left. + (-1)^{N+1} \ket{\eta}_{1}\ket{\eta}_{2}\ldots\ket{\eta}_{N} \right] ,
\end{align}
where the factor $N!$ is a result of the $N!$ possibilities for how the $N$ photons can transit from input modes to output modes.
This proves that GHZ states can be generated from letting $N$ independent photons interfere in a nonsymmetric $2N$-port network for arbitrary $N$. The success probability $P_{2N,\text{ps}}$ for such an event to occur calculates to [see Eq.~(\ref{eq:P_ps})]
\begin{align}
	P_{2N,\text{ps}} = \frac{(N!)^{2}}{2^{N-1}N^{2N}} .
\label{eq:P2N}
\end{align}
We plot $P_{2N,\text{ps}}$ in Fig.~\ref{fig:Probability} to compare it to our solutions using SMSs for even and odd $N$, which we discuss in the two following subsections.

\subsection{Symmetric N-port Splitters (odd N)}
\label{sec:MultiModeGHZ}

Next, let us discuss the generation of $N$-mode $N$-photon GHZ states using SMSs [see Eq.~(\ref{eq:Fourier})] and $N$-photon post-selection. For this, we will investigate $N$-photon input states, where each input mode contains exactly one photon (see Fig.~\ref{fig:SetupNetwork}), and show that this leads to a solution for odd $N$.

We start with the $N$-photon input state
\begin{align}
	&\ket{\Psi_{N,\text{in}}} = \frac{1}{\sqrt{2}^{N}}\prod_{k=1}^{N} \left[ e^{-\text{i}\theta_{k}} a_{\mu,k}^{\dagger} + e^{\text{i}\theta_{k}}a_{\eta,k}^{\dagger} \right] \ket{0} .
	\label{eq:InputState}
\end{align}
with $\theta_{k} = (k-1) 2\pi/N$, $k=1,2,\ldots,N$. In comparison to the input state given in Eq.~(\ref{eq:InputState2}) this state has an additional phase term, which leads to different output states for even $N$, in particular a GHZ state for $N=4$.

Employing the transformation relation of the SMS [see Eq.~(\ref{eq:TransformationRelationDFT})] we obtain the output state
\begin{align}
\label{eq:OutputState}
	\ket{\Psi_{N,\text{out}}} 
	 = & \frac{1}{\sqrt{2N}^{N}} \prod_{k=1}^{N} \left[ e^{-\text{i}\theta_{k}} \sum_{l=1}^{N} \omega_{N}^{(k-1)(l-1)} b_{\mu,l}^{\dagger} \right. \nonumber \\
	&  +  \left. e^{\text{i}\theta_{k}}\sum_{l=1}^{N} \omega_{N}^{(k-1)(l-1)} b_{\eta,l}^{\dagger} \right] \ket{0} . 
\end{align}

Let us now investigate the explicit form of the output state when post-selecting for $N$-mode coincidences, i.e., each output mode being occupied by a single photon.
To know how the post-selected state looks like, we need to know which output distributions, i.e., combinations of photons in the states $\ket{\mu}$ and $\ket{\eta}$, remain. For this we calculate the success probability of each output state that contains exactly one photon per output mode [see Eq.~(\ref{eq:P_ps})] and go through all combinations of $\ket{\mu}$ and $\ket{\eta}$.

We start by dividing the complete set (totally ordered) of occupied output modes $\sigma=\{1,2,\ldots,N\}$ into two subsets according to the internal states of the output photons, i.e., $\ket{\mu}$ and $\ket{\eta}$, each (see example in Fig.~\ref{fig:PhotonDistributions}).
For $N_{1}$ output photons being in the state $\ket{\mu}$ and $N_{2}$ output photons being in the state $\ket{\eta}$, with $N=N_{1}+N_{2}$, the two subsets are given by $\sigma_{\mu}=\{m_{1},m_{2},\ldots,m_{N_{1}}\}$ and $\sigma_{\eta}=\{h_{1},h_{2},\ldots,h_{N_{2}}\}$, with $\sigma=\sigma_{\mu} \cup \sigma_{\eta}$. Here, $m_{i}$ ($h_{i}$) denotes the output mode of the $i$th photon in the state $\ket{\mu}$ ($\ket{\eta}$). The probability to detect such a distribution is determined by the term [see Eq.~(\ref{eq:OutputState})]
\begin{align}
	&	\sum_{\mathcal{P}_{\delta}}  \left[ \prod_{l=1}^{N_{1}}  \omega_{N}^{(m_{l}-1)(\delta(l)-1)} e^{-\text{i}\theta_{\delta(l)}} \right] \nonumber \\
	&  \times \left[ \prod_{l'=1}^{N_{2}}  \omega_{N}^{(h_{l'}-1)(\delta(N_{1}+l')-1)} e^{\text{i}\theta_{\delta(N_{1}+l')}} \right] \nonumber \\
	& =	\sum_{\mathcal{P}_{\delta}} \left[ \prod_{l=1}^{N_{1}} \omega_{N}^{(m_{l}-2)(\delta(l)-1)} \right] \left[ \prod_{l'=1}^{N_{2}} \omega_{N}^{h_{l'}(\delta(N_{1}+l')-1)}  \right] ,
\label{eq:PostselectHV}
\end{align}
where we sum over the permutations $\mathcal{P}_{\delta}$ of the set of input modes $\delta=\{1,2,\ldots,N\}$, while the products are running over all output modes. In total this corresponds to summing over all possible photon input configurations that can contribute to a specific output state. In line 3 of Eq.~(\ref{eq:PostselectHV}) we made use of the definitions of $\omega_{N}=\exp(\text{i} 2\pi/N)$ and $\theta_{k}=(k-1)2\pi/N$ to combine the phases of the input photons and the phases of the multiport. This effectively reduces every $m_{l}$ by \textit{one} and, at the same time, increases every $h_{l'}$ by \textit{one}.

Instead of permuting the input modes $\delta=\{1,2,\ldots,N\}$ and keep the effective output modes $\sigma'=\{m_{1}-1,\ldots,m_{N_{1}}-1,h_{1}+1,\ldots,h_{N_{2}}+1\}$ fixed, one can also permute the effective output modes and keep the input modes fixed to sum over all possible combinations.
This allows us to rewrite Eq.~(\ref{eq:PostselectHV}) in the form
\begin{align}
\sum_{\mathcal{P}_{\sigma'}} \prod_{l=1}^{N} \omega_{N}^{(l-1)(\sigma'(l)-1)} .
\label{eq:PostselectHV2}
\end{align}
Since $\sigma'$ contains the elements $1,2,\ldots,N$, with $N_{1}$ elements being reduced by \textit{one} and $N_{2}$ elements being increased by \textit{one} we can apply the ZTL [see Eq.~(\ref{eq:ZTL})] and find the following condition for non-zero output distributions:
\begin{align}
	\sum_{l=1}^{N} \sigma'(l) & = \sum_{l=1}^{N} \sigma(l) - N_{1} + N_{2} \nonumber \\
	& = \frac{(N+1)N}{2} - N + 2 N_{2} \stackrel{!}{=} 0 \ \text{mod}(N)  ,
\label{eq:SigmaConditionHV}
\end{align}
where we made use of $N=N_{1}+N_{2}$.

Testing the condition given in Eq.~(\ref{eq:SigmaConditionHV}) by dividing by $N$, we obtain for odd $N$ ($\equiv N_{\text{o}}$)
\begin{align}
	\frac{(N_{\text{o}}+1)}{2} - 1 + 2 \frac{N_{2}}{N_{\text{o}}} \stackrel{?}{=} \text{integer} , \text{with} \ N_{2}=0,1,\ldots,N_{\text{o}} \, ,
\end{align}
where the first two terms are always integers.
Since $N_{2}\neq N_{\text{o}}/2$ for all $N_{\text{o}}$, the only possible constellations that can be measured coincidentally are all photons in the state $\ket{\mu}$, i.e., $N_{2}=0$, or all photons in the state $\ket{\eta}$, i.e., $N_{2}=N_{\text{o}}$. This means that $\sigma'$ either contains the elements $0,1,\ldots,N-1$ or $2,3,\ldots,N+1$. As $\omega_{N}^{x}=\omega_{N}^{x+N}$ ($x \in \mathbb{Z}$), Eq.~(\ref{eq:PostselectHV2}) is identical in both cases, and the resulting state can only be of the form of a GHZ-state. Note that Eq.~(\ref{eq:PostselectHV2}), identical to boson sampling with indistinguishable bosons, corresponds to the permanent of a matrix and is known to be hard to compute~\cite{Aaronson2011}.
Therefore, we calculate the success probabilities [see Eq.~(\ref{eq:P_ps})] for up to $N=11$ numerically and plot the results in Fig.~\ref{fig:Probability}.

Having found a working solution for odd $N$, we now proceed to investigate the possible output distributions for even $N$ ($\equiv N_{e}$). From  Eq.~(\ref{eq:SigmaConditionHV}) we find
\begin{align}
	 N_{2}  \stackrel{?}{=} \frac{N_{e}(2\! \times \! \text{integer} + 1)}{4}  , \text{with} \ N_{2}=0,1,\ldots,N_{\text{e}} \, .
\end{align}
This condition can only be fulfilled if $N_{\text{e}}$ is a multiple of $4$ and we find possible output states for $N_{2}=N_{\text{e}}/4$ and $N_{2}=3N_{\text{e}}/4$. However, this result is identical to a GHZ state only in the special case of $N=4$.

If $N_{\text{e}}$ is not a multiple of $4$, the condition cannot be fulfilled and all $N$-photon coincidences are suppressed. This behavior demonstrates a totally destructive HOM interference for partially distinguishable photons and coincides with the extended suppression law formulated in Ref.~\cite{Dittel2018}, where the authors discuss cyclic input distributions of partially distinguishable photons.

Hence, we have shown that sending an input state of the form of Eq.~(\ref{eq:InputState}) into a SMS generates an $N$-mode GHZ state for odd $N$. For even $N$ though, this scheme does not lead to the generation of GHZ states (exception: $N=4$) due to the generalized HOM effect. 
Note that one could instead utilize the input state given by Eq.~(\ref{eq:InputState2}). Similarly, as before, this would generate GHZ states for odd $N$ with the same probability, but it would not generate GHZ states for even $N$ (exception: $N=2$). However, post-selecting for $N$-photon coincidences in a single output mode, would in this case lead to single-mode GHZ states, which could subsequently be split up into $N$ spatial modes. We discuss this in Appendix~\ref{sec:SingleModeGHZ} and show that the overall success probability becomes identical to the one derived for $2N$-networks [see Eq.~(\ref{eq:P2N})].

In the next subsection we present an adjusted scheme to generate $N$-mode GHZ states for even $N$.

\subsection{Symmetric N-port Splitters (even N)}
\label{sec:MultiModeGHZEven}

\begin{figure}[t]
	\centering
		\includegraphics[width=0.95\columnwidth]{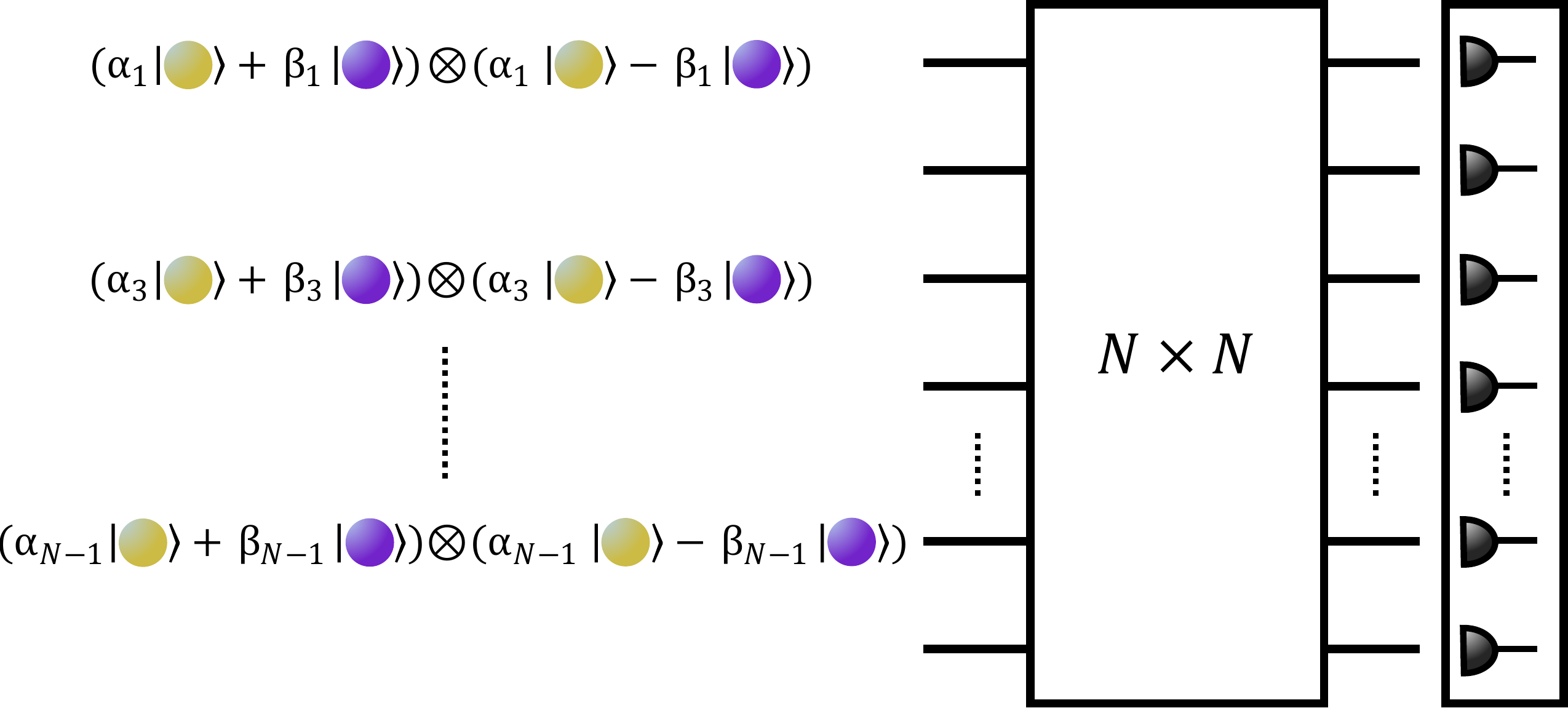}
	\caption{Unitary network used to generate GHZ states for even $N$. Two perpendicular photons are injected into every second input mode. Post-selection is used to detect $N$-mode $N$-photon coincidences at the $N$ output modes of the network.}
	\label{fig:SetupNetwork2}
\end{figure}

To generate a GHZ state with an even number of photons, we inject multiple photons into the same input mode.
We adjust the input state from Eq.~(\ref{eq:InputState2}) so that there are always two perpendicular photons entering the same input mode whereby we only use every second input mode (see Fig.~\ref{fig:SetupNetwork2}). Due to the symmetry of the setup, it is not important if we start with the first or the second input mode. We choose to start with the first one, i.e., $k=2\kappa-1$ ($\kappa=1,2,\ldots,N/2$). This reduces the DFT of the SMS to [see Eq.~(\ref{eq:Fourier})]
\begin{align}
	U_{k l} = \frac{1}{\sqrt{N}} \omega_{N}^{(2\kappa-2)(l-1)} = \frac{1}{\sqrt{N}} \tilde{\omega}_{N}^{(\kappa-1)(l-1)} ,
\label{eq:Fourier2}
\end{align}
with $\tilde{\omega}=\exp[\text{i} 2\pi/(N/2)]$ being the $N/2$th root of unity. In contrast to before, the conditions stemming from the ZTL now depend on mod$(N/2)$ instead of mod$(N)$.

The adjusted input state for $N_{e}$, i.e., even $N$, reads
\begin{align}
	\ket{\Psi_{N_{\text{e}},\text{in}}} = & \frac{1}{\sqrt{2}^{N}}\left\{ \prod_{\kappa=1}^{N/2}\left( a_{\mu,k}^{\dagger} + e^{\text{i}\theta_{\kappa}}a_{\eta,k}^{\dagger} \right) \right. \nonumber \\
	& \left. \vphantom{\prod_{\kappa=1}^{N/2}} \times \left( a_{\mu,k}^{\dagger} - e^{\text{i}\theta_{\kappa}}a_{\eta,k}^{\dagger} \right) \right\} \ket{0} ,
\label{eq:InputStateEven}
\end{align}
which simplifies to 
\begin{align}
	&\ket{\Psi_{N_{\text{e}},\text{in}}} =\nonumber \\
	& \frac{1}{\sqrt{2}^{N}} \left\{ \prod_{\kappa=1}^{N/2}\left[ \left( a_{\mu,k}^{\dagger} \right)^{2} -  e^{\text{i} 2 \theta_{\kappa}} \left( a_{\eta,k}^{\dagger}\right)^{2} \right] \vphantom{\prod_{k=1}^{N/2}} \right\} \ket{0} .
\label{eq:InputStateEven2}
\end{align}
From Eq.~(\ref{eq:InputStateEven2}) one sees that each of the $N/2$ input modes 
can be written as either two photons in $\ket{\mu}$ or $\ket{\eta}$, respectively.
Hence, only output distributions of even numbers of $\ket{\mu}$ and $\ket{\eta}$ can arise. Using the transformation relation given by Eq.~(\ref{eq:Fourier2}) yields the complete output state
\begin{align}
	\ket{\Psi_{N_{\text{e}},\text{out}}} & = \frac{1}{\sqrt{2}^{N}\sqrt{N}^{N}} \left\{  \prod_{\kappa=1}^{N/2}\left[ \left( \sum_{l=1}^{N} \tilde{\omega}^{(\kappa-1)(l-1)} b_{\mu,l}^{\dagger} \right)^{2}  \right. \right. \nonumber \\
	&\phantom{{}={}} \left. \left. - e^{\text{i} 2 \theta_{\kappa}} \left( \sum_{l=1}^{N} \tilde{\omega}^{(\kappa-1)(l-1)} b_{\eta,l}^{\dagger} \right)^{2} \right] \vphantom{\prod_{\kappa=1}^{N/2}} \right\}  \ket{0} .
\label{eq:OutputStateEven}
\end{align}

Identically to the case of odd-$N$ GHZ states,  we want to know the explicit form of the output state when post-selecting for $N$-mode coincidences, i.e., each output mode being occupied by exactly one photon.
Therefore, we need to calculate the success probability of each output distribution [see Eq.~(\ref{eq:P_ps})], i.e., each combination of $N=N_{1}+N_{2}$ photons in the states $\ket{\mu}$ and $\ket{\eta}$, respectively, where now $N_{1}$, $N_{2}$ and $N$ have to be even.

The complete set of output modes $\sigma=\{1,2,\ldots,N\}$ is, again, divided into the two totally ordered subsets $\sigma_{\mu}=\{m_{1},m_{2},\ldots,m_{N_{1}}\}$ and $\sigma_{\eta}=\{h_{1},h_{2},\ldots,h_{N_{2}}\}$, with $\sigma=\sigma_{\mu}\cup \sigma_{\eta}$  (see example in Fig.~\ref{fig:PhotonDistributions}). The output mode of the $i$th output photon in the state $\ket{\mu}$ ($\ket{\eta}$) is hence given by $m_{i}$ ($h_{i}$). 
We can now explicitly discuss the different input terms given by Eq.~(\ref{eq:InputStateEven2}), which lead to $\sigma_{\mu}$ and $\sigma_{\eta}$.

Since there are always two identical photons originating from the same input,
we can divide all photons into groups of two and then go through all possible pairings. W.l.o.g., we can combine $m_{1}$ and $m_{2}$, $m_{3}$ and $m_{4}$, \ldots, $m_{N_{1}-1}$ and $m_{N_{1}}$, and define $M_{1}=m_{1}+m_{2}$, \ldots, $M_{N_{1}/2}=m_{N_{1}-1}+m_{N_{1}}$ for the $N_{1}$ photons in the state $\ket{\mu}$ (analogously we define $H_{1}=h_{1}+h_{2}$, \ldots, $H_{N_{1}/2}=h_{N_{2}-1}+h_{N_{2}}$ for the $N_{2}$ photons in the state $\ket{\eta}$). The probability of such an output term is determined by
\begin{align}
	&	\sum_{\mathcal{P}_{\tilde{\delta}}} \left[ \prod_{l=1}^{N_{1}/2} \tilde{\omega}_{N}^{(M_{l}-2) (\tilde{\delta}(l)-1)}  \right] \nonumber \\
	& \times \left[ \prod_{l'=1}^{N_{2}/2} \tilde{\omega}_{N}^{(H_{l'}-2) (\tilde{\delta}(N_{1}/2+l')-1)} e^{\text{i} 2 \theta_{\tilde{\delta}(N_{1}/2+l')}} \right]   \nonumber \\
	& = 	\sum_{\mathcal{P}_{\tilde{\delta}}} \! \left[\! \prod_{l=1}^{N_{1}/2} \! \tilde{\omega}_{N}^{(M_{l}-2) (\tilde{\delta}(l)-1)} \! \right] \! \left[\! \prod_{l'=1}^{N_{2}/2} \! \tilde{\omega}_{N}^{(H_{l'}-1) (\tilde{\delta}(N_{1}/2+l')-1)} \!  \right] \! ,
\label{eq:PostselectHVEven}
\end{align}
where we sum over the permutations $\mathcal{P}_{\tilde{\delta}}$ of every second input mode $\tilde{\delta}=\{1,2,\ldots,N/2\}$ (with the definition of $\tilde{\omega}_{N}$ it becomes $2\tilde{\delta}(l)-1$ for every $l$, i.e., $\{1,3,\ldots,N-1\}$). In line 3 of Eq.~(\ref{eq:PostselectHVEven}) we made use of the definitions of $\tilde{\omega}_{N}=\exp[\text{i} 2\pi/(N/2)]$ and $\theta_{k}=(k-1)2\pi/N$ to combine the phases of the input photons and the phases of the MS. The argument is now identical to the one presented for odd $N$ in Sec.~\ref{sec:MultiModeGHZ}, and we can also sum over the permutation of adjusted output modes $\tilde{\sigma}'=\{M_{1},M_{2},\ldots,M_{N_{1}/2},H_{1}+1,H_{2}+1,\ldots,H_{N_{2}/2}+1\}$, while keeping the input modes fixed, i.e.,
\begin{align}
\sum_{\mathcal{P}_{\tilde{\sigma}'}} \prod_{l=1}^{N/2} \tilde{\omega}_{N}^{(l-1)(\tilde{\sigma}'(l)-2)} .
\label{eq:PostselectPM}
\end{align}
To have a nonzero contribution we need to sum over all $M_{i}$ and $H_{j}$ and add an additional $N_{2}/2$, leading to the condition [see ZTL in Eq.~(\ref{eq:ZTL})]
\begin{align}
	& \sum_{i=1}^{N_{1}/2} M_{i} + \sum_{j=1}^{N_{2}/2} H_{j} + \frac{N_{2}}{2} \nonumber \\
	& = \frac{N(N+1)}{2} + \frac{N_{2}}{2}\stackrel{!}{=} 0 \ \text{mod}(\frac{N}{2}) .
\end{align}
To fulfill this condition, $N_{2}$ has to be equal to $0$ or $N$, whereby $\tilde{\sigma}'=\{1,2,\ldots,N/2\}$ or $\{2,3,\ldots,N/2+1\}$. Since Eq.~(\ref{eq:PostselectPM}) is identical in both cases ($\tilde{\omega}_{N}^{x}=\tilde{\omega}_{N}^{x+N/2}$ for $x \in \mathbb{Z}$) this confirms the generation of an $N$-mode $N$-photon GHZ state for every even $N$. As in the case of odd $N$, there is no analytical solution for the success probability. We have numerically calculated it for $N=2,4,\ldots,12$ and plotted the values in Fig.~\ref{fig:Probability}.

\section{Comparison with other linear-optical methods for GHZ state generation}
\label{sec:Discussion}

\begin{figure}[t]
	\centering
		\includegraphics[width=0.8\columnwidth]{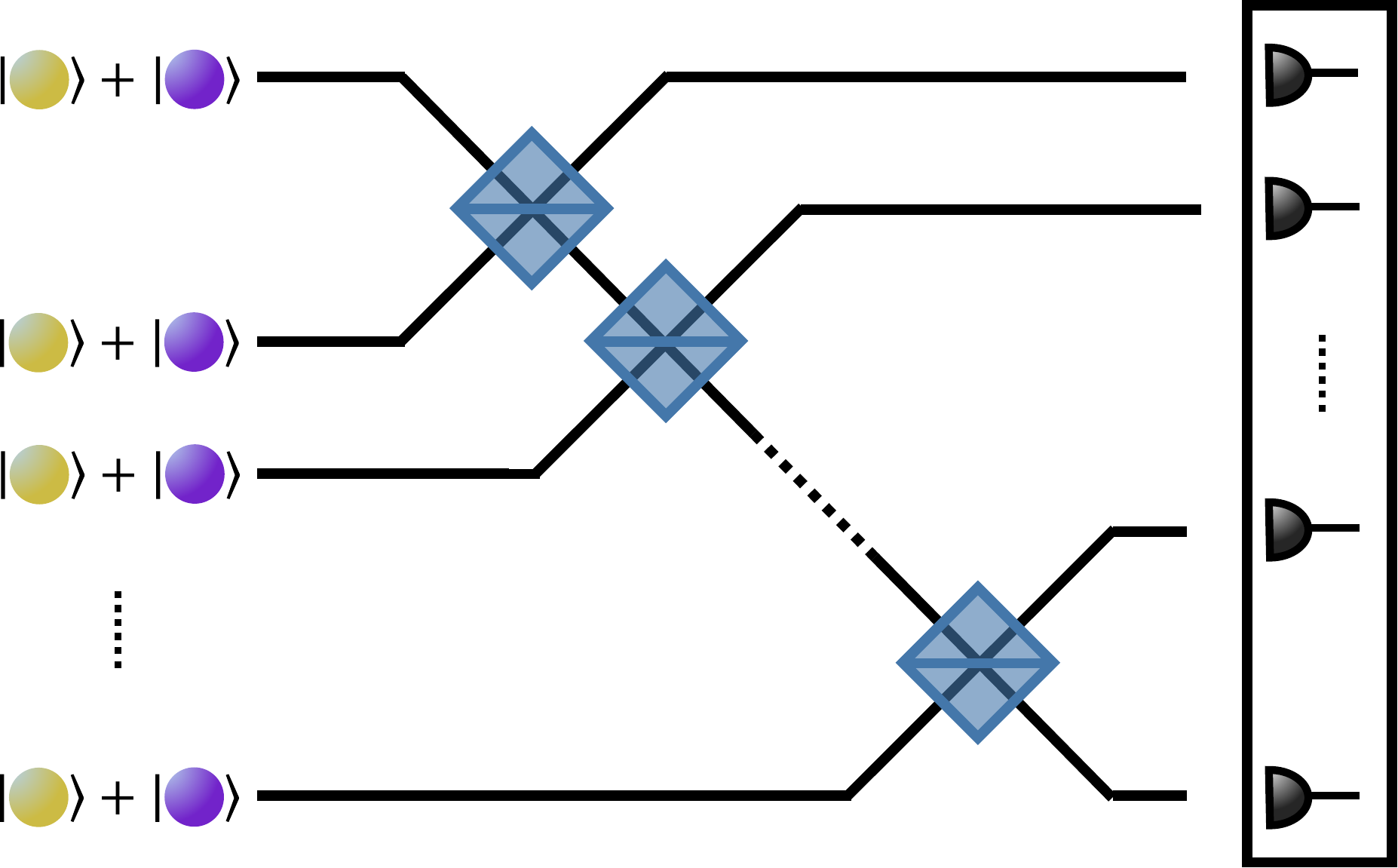}
	\caption{Setup of the standard scheme to post-select $N$-mode polarization-encoded GHZ states starting from $N$ independent photons, each in the polarization state $\ket{+}$. The $N$ photons interfere at $N-1$ polarizing beam splitters (blue squares) and are finally coincidentally detected at $N$ detectors. The figure has been adapted from Ref.~\cite{Sagi2003}.}
	\label{fig:SetupStandard}
\end{figure}

\begin{figure}[t]
	\centering
		\includegraphics[width=1\columnwidth]{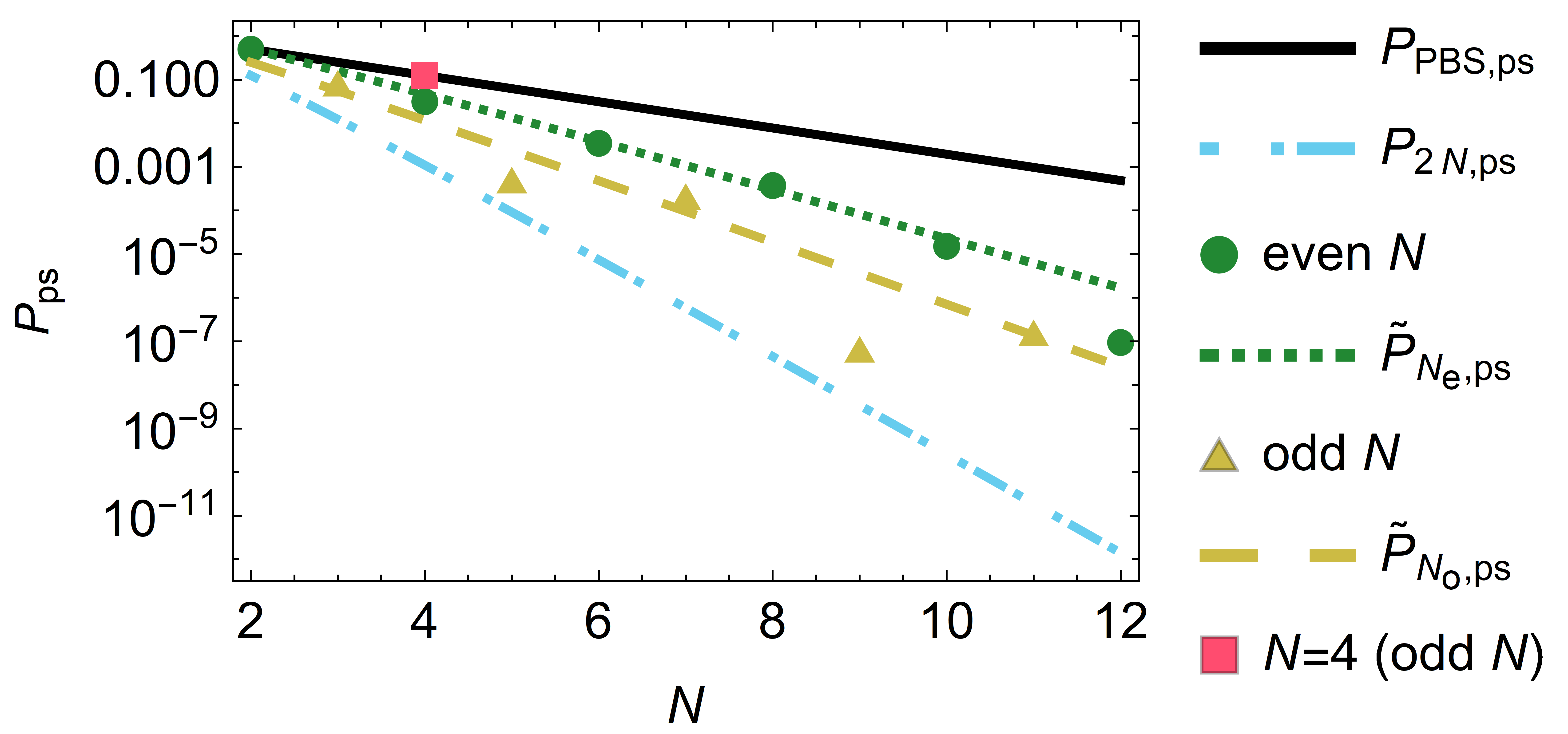}
	\caption{Plot of the different success probabilities. The dotted-dashed (light blue) line shows the success probability $P_{2N,\text{ps}}$ for a $2N$-port nonsymmetric network as described in Sec.~\ref{sec:2NNetwork}. The triangles (gold) and the circles (green) show the numerically calculated success probabilities for odd $N$ (Sec.~\ref{sec:MultiModeGHZ}) and even $N$ (Sec.~\ref{sec:MultiModeGHZEven}), respectively, while the dashed (gold) and dotted (green) lines display the roughly estimated behavior of the two different methods, i.e., $\tilde{P}_{N_{\text{o}},\text{ps}}$ and $\tilde{P}_{N_{\text{e}},\text{ps}}$. As a comparison, we plot the success probabilities of the standard method $P_{\text{PBS},\text{ps}}$ [solid (black) line] given by Eq.~(\ref{eq:SuccessProbPBS}).
	The red square marks the probability using $N=4$ photons in a symmetric BS with the input configuration $\ket{+-+-}$ \cite{Lim2005}, which coincides with the presented method for odd $N$.}
	\label{fig:Probability}
\end{figure}

In this section, we compare the three different methods to generate $N$-mode $N$-photon GHZ states discussed in this paper with respect to their success probabilities.
We further compare them to one of the standard methods for post-selecting polarization-encoded GHZ states~\cite{Sagi2003}, which we briefly introduce here. In this method one starts with $N$ independent photons, each in the polarization state $\ket{+}$, which then interfere at $N-1$ polarizing beam splitters (PBSs) (see Fig.~\ref{fig:SetupStandard}).
In combination with post-selecting one photon per output mode, this scheme lets one generate polarization-encoded $N$-photon GHZ states with a success probability of~\cite{Sagi2003}
\begin{align}
	P_{\text{PBS,ps}} = \frac{1}{2^{N-1}} .
\label{eq:SuccessProbPBS}
\end{align}
This result follows from the fact that the method using PBSs generates a total of $2^{N}$ possible output distributions. Two of them fulfill the post-selection condition and belong to the GHZ state. We plot this probability in Fig.~\ref{fig:Probability} together with the success probabilities of all three schemes presented above. It can be seen from the plot that for the SMS method in the cases $N=2$ (strategy for even $N$, Sec.~\ref{sec:MultiModeGHZEven}) and $N=4$ (strategy for odd N, Sec.~\ref{sec:MultiModeGHZ}) the probabilities are identical to the PBS method [see Eq.~(\ref{eq:SuccessProbPBS})]. For any other $N$ the SMS method produces lower success probabilities than the method using PBSs.
However, the SMS method has the advantage that it offers a better flexibility. It works for different choices of internal degrees of freedom and not only polarization like the PBS method (see Appendix~\ref{sec:InternalDoF}), and, importantly, it can produce different entangled output states, e.g., the $N$-mode W state~\cite{Lim2005}, without the need for changing the experimental setup. The latter has been demonstrated in Ref.~\cite{Kumar2022}, where for $N=3$ photons tripartite states of the GHZ and W class have been generated by adjusting the input polarization of the photons.

Intuitively, the scaling of the SMS method can be understood from roughly estimating the number of all possible outputs and the number of outputs which contribute to the GHZ state. Injecting $N$ photons according to the methods described above into an $N$-port MS gives a total of $2^{N}$ possible inputs for odd $N$ [see Eq.~(\ref{eq:InputState})] and $2^{N/2}$ for even $N$ [see Eq.~(\ref{eq:InputStateEven2})]. Each input can then generate a maximal number of $N^{N}$ outputs since each of the $N$ photons can go to any of the $N$ output modes. For each GHZ state there are $2\times N!$ valid output possibilities, which denote the two cases where all photons are identical and the $N!$ possible transitions of the $N$ input photons to the $N$ output modes. This leads to the estimated scalings for odd $N$
\begin{align}
\tilde{P}_{N_{\text{o}},\text{ps}} = \frac{N!}{2^{N-1}N^{N}},
\label{eq:PNo}
\end{align}
and even $N$
\begin{align}
\tilde{P}_{N_{\text{e}},\text{ps}} = \frac{N!}{2^{N/2-1}N^{N}},
\label{eq:PNe}
\end{align}
which we plot together with the numerically calculated results for up to $N=12$ in Fig.~\ref{fig:Probability}.

In comparison to the $2N$-port method presented in Sec.~\ref{sec:2NNetwork}, which can also produce output states of different entanglement classes, the success probabilities with respect to the methods making use of the SMS (Secs.~\ref{sec:MultiModeGHZ} and \ref{sec:MultiModeGHZEven}) are higher. In Appendix~\ref{sec:SingleModeGHZ} we show that, additionally, single-mode $N$-photon GHZ states can be generated with the help of SMSs, with a success probability that is identical to Eq.~(\ref{eq:PNo}) [see Eq.~(\ref{eq:PSingleMode})].

\section{Conclusion}
\label{sec:Conclusion}

In conclusion, we have shown that $N$-photon $N$-mode GHZ states can be generated using interference in SMSs. We have presented novel schemes for even and odd $N$ demonstrating that the GHZ state generation is possible for arbitrary $N$.
In particular, by making use of the zero transmission law \cite{Tichy2010}, we have shown that when sending in partially distinguishable photons in specific input states, only terms describing a GHZ state can occur.
Since the success probability is described by a term identical to the permanent of a random matrix, it cannot be calculated analytically. We have presented estimations of the scaling and numerical results for up to $N=12$. To our knowledge, this generalization has been missing so far. In contrast to the standard method discussed in Sec.~\ref{sec:Discussion}, which relies on polarization, the presented interferences of independent photons in SMSs are not limited to polarization but work for any internal degree of freedom capable of describing a photonic qubit (see Appendix~\ref{sec:InternalDoF}). Therefore, it opens up the possibility to use any symmetric $N$-port splitter as an $N$-partite entanglement generator producing different entangled states, e.g., GHZ and W states, for different internal degrees of freedom by simply adjusting the input states.

Depending on the goal, this flexibility can even compensate for lower success probabilities compared to schemes which can only generate a specific entangled state. For example, quantum networks would greatly benefit from a server that (possibly randomly) sends out different entangled states to multiple parties without the need for changing the experimental setup. This would easily allow for different quantum protocols~\cite{Murta2020}. Moreover, in combination with additional fusion gates, multiple SMSs could be used as resources for building cluster states and, therefore, photonic quantum computing~\cite{Gimeno-Segovia2015,Omkar2021}. The possibility of even generating states belonging to additional entanglement classes, e.g., cluster states or G states~\cite{Kumar2022}, by adjusting the input states would further enhance the range of possible protocols and fuseable quantum states.

The presented schemes can not only be used to post-select, but also herald the successful generation of $N$-partite entanglement.
In similarity to Ref.~\cite{Bastin2009}, the MSs can be used to entangle $N$ single-photon emitters each defined by a $\Lambda$ scheme, e.g., trapped ions, or
quantum dots. Since in this case one does not know the state of the emitted photons, the post-selection process would have to be extended in order to analyze the photons' internal degrees of freedom.
This also automatically allows for heralding entangled $N$-photon states, by starting with pairs of entangled photons instead of single photons.

Finally, it is an interesting question to study the different methods for GHZ-state generation under the influence of loss and experimental imperfections such as distinguishability and mixedness~\cite{Jones2022}. Furthermore, our method could be combined with approaches from Ref.~\cite{Meyer-Scott2022} which is using active feedforward and multiplexing to increase the generation probabilities.

\begin{acknowledgments}
We thank Jelena Mackeprang for fruitful discussions and helpful comments on the manuscript. 
We acknowledge support from the Carl Zeiss Foundation, the Center for Integrated Quantum Science and Technology (IQ$^\text{ST}$), the German Research Foundation (DFG), the Federal Ministry of Education and Research (BMBF, projects SiSiQ and PhotonQ), the Federal Ministry for Economic Affairs and Energy (BMWi, project PlanQK), and the Competence Center Quantum Computing Baden-W\"urttemberg (funded by the Ministerium f\"ur Wirtschaft, Arbeit und Tourismus Baden-W\"urttemberg, project QORA).
\end{acknowledgments}

\appendix

\section{Single-mode GHZ states: Symmetric Multiport Splitters}
\label{sec:SingleModeGHZ}

To find a solution working for any $N$ and SMSs we make use of an idea presented in Ref.~\cite{Hofmann2004} which utilizes a linear array of two-mode beam splitters.
This scheme generates GHZ states for arbitrary $N$. However, it requires the post-selection of $N$-photon coincidences in a single output mode. In analogy to the original idea we use the input state given by Eq.~(\ref{eq:InputState2}) and send it into an SMS.
Note that the authors of Ref.~\cite{Seron2022} also investigate this particular state as an input to SMSs, however, in the context of multimode bosonic bunching.

The post-selected single-mode output state for the output mode $l$ takes the form
\begin{align}
	& \ket{\psi_{N_{l},\text{ps}}} \nonumber \\
	& = \frac{1}{\sqrt{2}^{N}} \left\{ \prod_{k=1}^{N} U_{kl} \left[ b_{\mu,l}^{\dagger} + \exp(\text{i}\theta_{k}) b_{\eta,l}^{\dagger} \right] \right\} \ket{0} \nonumber \\
	& = \frac{1}{\sqrt{2N}^{N}} \left\{ \prod_{k=1}^{N} \omega_{N}^{(k-1)(l-1)} \left[  b_{\mu,l}^{\dagger} + \exp(\text{i}\theta_{k})  b_{\eta,l}^{\dagger} \right] \right\} \ket{0} , 
\end{align}
where the phase induced by the MS factors out and can be neglected. By choosing a fixed number of $N_{1}$ photons in the state $\ket{\mu}$ and $N_{2}$ photons in the state $\ket{\eta}$ one arrives at the same conclusion as presented in Sec.~\ref{sec:2NNetwork}, namely that all terms except those of only identical photons are suppressed.
The final post-selected state is then of the form
\begin{align}
	\ket{\psi_{N_{l},\text{ps}}} & = \frac{1}{\sqrt{2N}^{N}} \left[ \left(b_{\mu,l}^{\dagger}\right)^{N} + (-1)^{N+1} \left(b_{\eta,l}^{\dagger}\right)^{N} \right] \ket{0} .
\end{align}
The respective success probability yields
\begin{align}
	P_{N_{l},\text{ps}} = \frac{N!}{2^{N-1}N^{N}} ,
\label{eq:PSingleMode}
\end{align}
which is identical to the success probability calculated for the original setup \cite{Hofmann2004}. By using a second $N$-port BS to transform the single-mode GHZ state into an $N$-mode GHZ state an additional factor of $N!/N^{N}$ is added to Eq.~(\ref{eq:PSingleMode}), which makes it identical to the result of the $2N$-network discussed above [see Eq.~(\ref{eq:P2N})].

Note that, when using the state given in Eq.~(\ref{eq:InputState}) as an input state, in the case of even $N$, the single-mode output state is not a GHZ state anymore. Then, the output state also contains a term with $N/2$ photons in the state $\ket{\mu}$ and $N/2$ photons in the state $\ket{\eta}$, which for $N=4$ is identical to the Holland-Burnett state.

\section{Internal Degrees of Freedom}
\label{sec:InternalDoF}

The state of a single photon $\ket{\Psi_{k}}=\ket{F_{1},F_{2},\ldots,F_{n}}_{k}$ is determined by its internal degrees of freedom $F_{i}$ ($i=1,2,\ldots,n$) and its external degree of freedom $k$. In this work the external degree of freedom $k=1,2,\ldots,N$ denotes the respective input or output mode of the MS in which the photon enters or exits the MS (see Fig.~\ref{fig:NPort_basic}). The internal degrees of freedom denote the photons' properties which can make the photons distinguishable, e.g., polarization or wavelength, and lead to many-particle interference in the external degree of freedom~\cite{Minke2021}. Let us assume that for all photons considered in this work all but one internal degree of freedom (=property) $F$ are identical and we can write $\ket{\Psi_{k}}=\ket{F}_{k}$. The distinguishability of two photons is then given by~\cite{Menssen2017}
\begin{align}
	\braket{\Psi_{i}|\Psi_{j}}=r_{ij}e^{i \varphi_{ij}},
\end{align}
where $r_{ij} \in (0,1)$ and $\varphi_{ij} \in (0,2\pi)$. 
The value of $r_{ij}$ tells how distinguishable two photons are. For $r_{ij}=0$ the two photons are perfectly distinguishable, e.g., a horizontally and a vertically polarized photon, i.e., $\braket{H|V}=0$. For $r_{ij}=1$ the two photons are identical, e.g., two identically polarized photons, i.e., $\braket{V|V}=1$.

To generate the entangled GHZ states as discussed in the main text one can utilize any internal degree of freedom that allows for preparing single photons in two orthogonal states. Examples are polarization ($\braket{H|V}=0$), wavelength ($\braket{\lambda_{1}|\lambda_{2}}=0$), orbital angular momentum ($\braket{l_{1}|l_{2}}=0$), and time-bin-encoding ($\braket{t_{1}|t_{2}}=0$).


\begin{thebibliography}{46}%
\makeatletter
\providecommand \@ifxundefined [1]{%
 \@ifx{#1\undefined}
}%
\providecommand \@ifnum [1]{%
 \ifnum #1\expandafter \@firstoftwo
 \else \expandafter \@secondoftwo
 \fi
}%
\providecommand \@ifx [1]{%
 \ifx #1\expandafter \@firstoftwo
 \else \expandafter \@secondoftwo
 \fi
}%
\providecommand \natexlab [1]{#1}%
\providecommand \enquote  [1]{``#1''}%
\providecommand \bibnamefont  [1]{#1}%
\providecommand \bibfnamefont [1]{#1}%
\providecommand \citenamefont [1]{#1}%
\providecommand \href@noop [0]{\@secondoftwo}%
\providecommand \href [0]{\begingroup \@sanitize@url \@href}%
\providecommand \@href[1]{\@@startlink{#1}\@@href}%
\providecommand \@@href[1]{\endgroup#1\@@endlink}%
\providecommand \@sanitize@url [0]{\catcode `\\12\catcode `\$12\catcode
  `\&12\catcode `\#12\catcode `\^12\catcode `\_12\catcode `\%12\relax}%
\providecommand \@@startlink[1]{}%
\providecommand \@@endlink[0]{}%
\providecommand \url  [0]{\begingroup\@sanitize@url \@url }%
\providecommand \@url [1]{\endgroup\@href {#1}{\urlprefix }}%
\providecommand \urlprefix  [0]{URL }%
\providecommand \Eprint [0]{\href }%
\providecommand \doibase [0]{https://doi.org/}%
\providecommand \selectlanguage [0]{\@gobble}%
\providecommand \bibinfo  [0]{\@secondoftwo}%
\providecommand \bibfield  [0]{\@secondoftwo}%
\providecommand \translation [1]{[#1]}%
\providecommand \BibitemOpen [0]{}%
\providecommand \bibitemStop [0]{}%
\providecommand \bibitemNoStop [0]{.\EOS\space}%
\providecommand \EOS [0]{\spacefactor3000\relax}%
\providecommand \BibitemShut  [1]{\csname bibitem#1\endcsname}%
\let\auto@bib@innerbib\@empty
%</preamble>
\bibitem [{\citenamefont {Hillery}\ \emph {et~al.}(1999)\citenamefont
  {Hillery}, \citenamefont {Bu\ifmmode~\check{z}\else \v{z}\fi{}ek},\ and\
  \citenamefont {Berthiaume}}]{Hillery1999}%
  \BibitemOpen
  \bibfield  {author} {\bibinfo {author} {\bibfnamefont {M.}~\bibnamefont
  {Hillery}}, \bibinfo {author} {\bibfnamefont {V.}~\bibnamefont
  {Bu\ifmmode~\check{z}\else \v{z}\fi{}ek}},\ and\ \bibinfo {author}
  {\bibfnamefont {A.}~\bibnamefont {Berthiaume}},\ }\bibfield  {title}
  {\bibinfo {title} {Quantum secret sharing},\ }\href
  {https://doi.org/10.1103/PhysRevA.59.1829} {\bibfield  {journal} {\bibinfo
  {journal} {Phys. Rev. A}\ }\textbf {\bibinfo {volume} {59}},\ \bibinfo
  {pages} {1829} (\bibinfo {year} {1999})}\BibitemShut {NoStop}%
\bibitem [{\citenamefont {Epping}\ \emph {et~al.}(2017)\citenamefont {Epping},
  \citenamefont {Kampermann}, \citenamefont {Macchiavello},\ and\ \citenamefont
  {Bru{\ss}}}]{Epping2017}%
  \BibitemOpen
  \bibfield  {author} {\bibinfo {author} {\bibfnamefont {M.}~\bibnamefont
  {Epping}}, \bibinfo {author} {\bibfnamefont {H.}~\bibnamefont {Kampermann}},
  \bibinfo {author} {\bibfnamefont {C.}~\bibnamefont {Macchiavello}},\ and\
  \bibinfo {author} {\bibfnamefont {D.}~\bibnamefont {Bru{\ss}}},\ }\bibfield
  {title} {\bibinfo {title} {{Multi-partite entanglement can speed up quantum
  key distribution in networks}},\ }\href
  {https://doi.org/10.1088/1367-2630/aa8487} {\bibfield  {journal} {\bibinfo
  {journal} {New J. Phys.}\ }\textbf {\bibinfo {volume} {19}},\ \bibinfo
  {pages} {093012} (\bibinfo {year} {2017})}\BibitemShut {NoStop}%
\bibitem [{\citenamefont {Grasselli}\ \emph {et~al.}(2018)\citenamefont
  {Grasselli}, \citenamefont {Kampermann},\ and\ \citenamefont
  {Bru{\ss}}}]{Grasselli2018}%
  \BibitemOpen
  \bibfield  {author} {\bibinfo {author} {\bibfnamefont {F.}~\bibnamefont
  {Grasselli}}, \bibinfo {author} {\bibfnamefont {H.}~\bibnamefont
  {Kampermann}},\ and\ \bibinfo {author} {\bibfnamefont {D.}~\bibnamefont
  {Bru{\ss}}},\ }\bibfield  {title} {\bibinfo {title} {{Finite-key effects in
  multipartite quantum key distribution protocols}},\ }\href
  {https://doi.org/10.1088/1367-2630/aaec34} {\bibfield  {journal} {\bibinfo
  {journal} {New J. Phys.}\ }\textbf {\bibinfo {volume} {20}},\ \bibinfo
  {pages} {113014} (\bibinfo {year} {2018})}\BibitemShut {NoStop}%
\bibitem [{\citenamefont {Thomas}\ \emph {et~al.}(2022)\citenamefont {Thomas},
  \citenamefont {Ruscio}, \citenamefont {Morin},\ and\ \citenamefont
  {Rempe}}]{Thomas2022}%
  \BibitemOpen
  \bibfield  {author} {\bibinfo {author} {\bibfnamefont {P.}~\bibnamefont
  {Thomas}}, \bibinfo {author} {\bibfnamefont {L.}~\bibnamefont {Ruscio}},
  \bibinfo {author} {\bibfnamefont {O.}~\bibnamefont {Morin}},\ and\ \bibinfo
  {author} {\bibfnamefont {G.}~\bibnamefont {Rempe}},\ }\bibfield  {title}
  {\bibinfo {title} {{Efficient generation of entangled multiphoton graph
  states from a single atom}},\ }\href
  {https://doi.org/10.1038/s41586-022-04987-5} {\bibfield  {journal} {\bibinfo
  {journal} {Nature}\ }\textbf {\bibinfo {volume} {608}},\ \bibinfo {pages}
  {677} (\bibinfo {year} {2022})}\BibitemShut {NoStop}%
\bibitem [{\citenamefont {Wang}\ \emph {et~al.}(2018)\citenamefont {Wang},
  \citenamefont {Luo}, \citenamefont {Huang}, \citenamefont {Chen},
  \citenamefont {Su}, \citenamefont {Liu}, \citenamefont {Chen}, \citenamefont
  {Li}, \citenamefont {Fang}, \citenamefont {Jiang}, \citenamefont {Zhang},
  \citenamefont {Li}, \citenamefont {Liu}, \citenamefont {Lu},\ and\
  \citenamefont {Pan}}]{Wang2018}%
  \BibitemOpen
  \bibfield  {author} {\bibinfo {author} {\bibfnamefont {X.-L.}\ \bibnamefont
  {Wang}}, \bibinfo {author} {\bibfnamefont {Y.-H.}\ \bibnamefont {Luo}},
  \bibinfo {author} {\bibfnamefont {H.-L.}\ \bibnamefont {Huang}}, \bibinfo
  {author} {\bibfnamefont {M.-C.}\ \bibnamefont {Chen}}, \bibinfo {author}
  {\bibfnamefont {Z.-E.}\ \bibnamefont {Su}}, \bibinfo {author} {\bibfnamefont
  {C.}~\bibnamefont {Liu}}, \bibinfo {author} {\bibfnamefont {C.}~\bibnamefont
  {Chen}}, \bibinfo {author} {\bibfnamefont {W.}~\bibnamefont {Li}}, \bibinfo
  {author} {\bibfnamefont {Y.-Q.}\ \bibnamefont {Fang}}, \bibinfo {author}
  {\bibfnamefont {X.}~\bibnamefont {Jiang}}, \bibinfo {author} {\bibfnamefont
  {J.}~\bibnamefont {Zhang}}, \bibinfo {author} {\bibfnamefont
  {L.}~\bibnamefont {Li}}, \bibinfo {author} {\bibfnamefont {N.-L.}\
  \bibnamefont {Liu}}, \bibinfo {author} {\bibfnamefont {C.-Y.}\ \bibnamefont
  {Lu}},\ and\ \bibinfo {author} {\bibfnamefont {J.-W.}\ \bibnamefont {Pan}},\
  }\bibfield  {title} {\bibinfo {title} {{18-Qubit Entanglement with Six
  Photons' Three Degrees of Freedom}},\ }\href
  {https://doi.org/10.1103/PhysRevLett.120.260502} {\bibfield  {journal}
  {\bibinfo  {journal} {Phys. Rev. Lett.}\ }\textbf {\bibinfo {volume} {120}},\
  \bibinfo {pages} {260502} (\bibinfo {year} {2018})}\BibitemShut {NoStop}%
\bibitem [{\citenamefont {Omran}\ \emph {et~al.}(2019)\citenamefont {Omran},
  \citenamefont {Levine}, \citenamefont {Keesling}, \citenamefont {Semeghini},
  \citenamefont {Wang}, \citenamefont {Ebadi}, \citenamefont {Bernien},
  \citenamefont {Zibrov}, \citenamefont {Pichler}, \citenamefont {Choi},
  \citenamefont {Cui}, \citenamefont {Rossignolo}, \citenamefont {Rembold},
  \citenamefont {Montangero}, \citenamefont {Calarco}, \citenamefont {Endres},
  \citenamefont {Greiner}, \citenamefont
  {Vuleti{\ifmmode\acute{c}\else\'{c}\fi}},\ and\ \citenamefont
  {Lukin}}]{Omran2019}%
  \BibitemOpen
  \bibfield  {author} {\bibinfo {author} {\bibfnamefont {A.}~\bibnamefont
  {Omran}}, \bibinfo {author} {\bibfnamefont {H.}~\bibnamefont {Levine}},
  \bibinfo {author} {\bibfnamefont {A.}~\bibnamefont {Keesling}}, \bibinfo
  {author} {\bibfnamefont {G.}~\bibnamefont {Semeghini}}, \bibinfo {author}
  {\bibfnamefont {T.~T.}\ \bibnamefont {Wang}}, \bibinfo {author}
  {\bibfnamefont {S.}~\bibnamefont {Ebadi}}, \bibinfo {author} {\bibfnamefont
  {H.}~\bibnamefont {Bernien}}, \bibinfo {author} {\bibfnamefont {A.~S.}\
  \bibnamefont {Zibrov}}, \bibinfo {author} {\bibfnamefont {H.}~\bibnamefont
  {Pichler}}, \bibinfo {author} {\bibfnamefont {S.}~\bibnamefont {Choi}},
  \bibinfo {author} {\bibfnamefont {J.}~\bibnamefont {Cui}}, \bibinfo {author}
  {\bibfnamefont {M.}~\bibnamefont {Rossignolo}}, \bibinfo {author}
  {\bibfnamefont {P.}~\bibnamefont {Rembold}}, \bibinfo {author} {\bibfnamefont
  {S.}~\bibnamefont {Montangero}}, \bibinfo {author} {\bibfnamefont
  {T.}~\bibnamefont {Calarco}}, \bibinfo {author} {\bibfnamefont
  {M.}~\bibnamefont {Endres}}, \bibinfo {author} {\bibfnamefont
  {M.}~\bibnamefont {Greiner}}, \bibinfo {author} {\bibfnamefont
  {V.}~\bibnamefont {Vuleti{\ifmmode\acute{c}\else\'{c}\fi}}},\ and\ \bibinfo
  {author} {\bibfnamefont {M.~D.}\ \bibnamefont {Lukin}},\ }\bibfield  {title}
  {\bibinfo {title} {{Generation and manipulation of
  Schr{\ifmmode\ddot{o}\else\"{o}\fi}dinger cat states in Rydberg atom
  arrays}},\ }\href {https://doi.org/10.1126/science.aax9743} {\bibfield
  {journal} {\bibinfo  {journal} {Science}\ }\textbf {\bibinfo {volume}
  {365}},\ \bibinfo {pages} {570} (\bibinfo {year} {2019})}\BibitemShut
  {NoStop}%
\bibitem [{\citenamefont {Murta}\ \emph {et~al.}(2020)\citenamefont {Murta},
  \citenamefont {Grasselli}, \citenamefont {Kampermann},\ and\ \citenamefont
  {Bru{\ss}}}]{Murta2020}%
  \BibitemOpen
  \bibfield  {author} {\bibinfo {author} {\bibfnamefont {G.}~\bibnamefont
  {Murta}}, \bibinfo {author} {\bibfnamefont {F.}~\bibnamefont {Grasselli}},
  \bibinfo {author} {\bibfnamefont {H.}~\bibnamefont {Kampermann}},\ and\
  \bibinfo {author} {\bibfnamefont {D.}~\bibnamefont {Bru{\ss}}},\ }\bibfield
  {title} {\bibinfo {title} {{Quantum Conference Key Agreement: A Review}},\
  }\href {https://doi.org/10.1002/qute.202000025} {\bibfield  {journal}
  {\bibinfo  {journal} {Adv. Quantum Technol.}\ }\textbf {\bibinfo {volume}
  {3}},\ \bibinfo {pages} {2000025} (\bibinfo {year} {2020})}\BibitemShut
  {NoStop}%
\bibitem [{\citenamefont {Hahn}\ \emph {et~al.}(2020)\citenamefont {Hahn},
  \citenamefont {de~Jong},\ and\ \citenamefont {Pappa}}]{Hahn2020}%
  \BibitemOpen
  \bibfield  {author} {\bibinfo {author} {\bibfnamefont {F.}~\bibnamefont
  {Hahn}}, \bibinfo {author} {\bibfnamefont {J.}~\bibnamefont {de~Jong}},\ and\
  \bibinfo {author} {\bibfnamefont {A.}~\bibnamefont {Pappa}},\ }\bibfield
  {title} {\bibinfo {title} {Anonymous quantum conference key agreement},\
  }\href {https://doi.org/10.1103/PRXQuantum.1.020325} {\bibfield  {journal}
  {\bibinfo  {journal} {PRX Quantum}\ }\textbf {\bibinfo {volume} {1}},\
  \bibinfo {pages} {020325} (\bibinfo {year} {2020})}\BibitemShut {NoStop}%
\bibitem [{\citenamefont {Thalacker}\ \emph {et~al.}(2021)\citenamefont
  {Thalacker}, \citenamefont {Hahn}, \citenamefont {de~Jong}, \citenamefont
  {Pappa},\ and\ \citenamefont {Barz}}]{Thalacker2021}%
  \BibitemOpen
  \bibfield  {author} {\bibinfo {author} {\bibfnamefont {C.}~\bibnamefont
  {Thalacker}}, \bibinfo {author} {\bibfnamefont {F.}~\bibnamefont {Hahn}},
  \bibinfo {author} {\bibfnamefont {J.}~\bibnamefont {de~Jong}}, \bibinfo
  {author} {\bibfnamefont {A.}~\bibnamefont {Pappa}},\ and\ \bibinfo {author}
  {\bibfnamefont {S.}~\bibnamefont {Barz}},\ }\bibfield  {title} {\bibinfo
  {title} {{Anonymous and secret communication in quantum networks}},\ }\href
  {https://doi.org/10.1088/1367-2630/ac1808} {\bibfield  {journal} {\bibinfo
  {journal} {New J. Phys.}\ }\textbf {\bibinfo {volume} {23}},\ \bibinfo
  {pages} {083026} (\bibinfo {year} {2021})}\BibitemShut {NoStop}%
\bibitem [{\citenamefont {Pogorelov}\ \emph {et~al.}(2021)\citenamefont
  {Pogorelov}, \citenamefont {Feldker}, \citenamefont {Marciniak},
  \citenamefont {Postler}, \citenamefont {Jacob}, \citenamefont
  {Krieglsteiner}, \citenamefont {Podlesnic}, \citenamefont {Meth},
  \citenamefont {Negnevitsky}, \citenamefont {Stadler}, \citenamefont
  {H\"ofer}, \citenamefont {W\"achter}, \citenamefont {Lakhmanskiy},
  \citenamefont {Blatt}, \citenamefont {Schindler},\ and\ \citenamefont
  {Monz}}]{Pogorelov2021}%
  \BibitemOpen
  \bibfield  {author} {\bibinfo {author} {\bibfnamefont {I.}~\bibnamefont
  {Pogorelov}}, \bibinfo {author} {\bibfnamefont {T.}~\bibnamefont {Feldker}},
  \bibinfo {author} {\bibfnamefont {C.~D.}\ \bibnamefont {Marciniak}}, \bibinfo
  {author} {\bibfnamefont {L.}~\bibnamefont {Postler}}, \bibinfo {author}
  {\bibfnamefont {G.}~\bibnamefont {Jacob}}, \bibinfo {author} {\bibfnamefont
  {O.}~\bibnamefont {Krieglsteiner}}, \bibinfo {author} {\bibfnamefont
  {V.}~\bibnamefont {Podlesnic}}, \bibinfo {author} {\bibfnamefont
  {M.}~\bibnamefont {Meth}}, \bibinfo {author} {\bibfnamefont {V.}~\bibnamefont
  {Negnevitsky}}, \bibinfo {author} {\bibfnamefont {M.}~\bibnamefont
  {Stadler}}, \bibinfo {author} {\bibfnamefont {B.}~\bibnamefont {H\"ofer}},
  \bibinfo {author} {\bibfnamefont {C.}~\bibnamefont {W\"achter}}, \bibinfo
  {author} {\bibfnamefont {K.}~\bibnamefont {Lakhmanskiy}}, \bibinfo {author}
  {\bibfnamefont {R.}~\bibnamefont {Blatt}}, \bibinfo {author} {\bibfnamefont
  {P.}~\bibnamefont {Schindler}},\ and\ \bibinfo {author} {\bibfnamefont
  {T.}~\bibnamefont {Monz}},\ }\bibfield  {title} {\bibinfo {title} {Compact
  ion-trap quantum computing demonstrator},\ }\href
  {https://doi.org/10.1103/PRXQuantum.2.020343} {\bibfield  {journal} {\bibinfo
   {journal} {PRX Quantum}\ }\textbf {\bibinfo {volume} {2}},\ \bibinfo {pages}
  {020343} (\bibinfo {year} {2021})}\BibitemShut {NoStop}%
\bibitem [{\citenamefont {Mooney}\ \emph {et~al.}(2021)\citenamefont {Mooney},
  \citenamefont {White}, \citenamefont {Hill},\ and\ \citenamefont
  {Hollenberg}}]{Mooney2021}%
  \BibitemOpen
  \bibfield  {author} {\bibinfo {author} {\bibfnamefont {G.~J.}\ \bibnamefont
  {Mooney}}, \bibinfo {author} {\bibfnamefont {G.~A.~L.}\ \bibnamefont
  {White}}, \bibinfo {author} {\bibfnamefont {C.~D.}\ \bibnamefont {Hill}},\
  and\ \bibinfo {author} {\bibfnamefont {L.~C.~L.}\ \bibnamefont
  {Hollenberg}},\ }\bibfield  {title} {\bibinfo {title} {{Generation and
  verification of 27-qubit Greenberger-Horne-Zeilinger states in a
  superconducting quantum computer}},\ }\href
  {https://doi.org/10.1088/2399-6528/ac1df7} {\bibfield  {journal} {\bibinfo
  {journal} {J. Phys. Commun.}\ }\textbf {\bibinfo {volume} {5}},\ \bibinfo
  {pages} {095004} (\bibinfo {year} {2021})}\BibitemShut {NoStop}%
\bibitem [{\citenamefont {Meyer-Scott}\ \emph {et~al.}(2022)\citenamefont
  {Meyer-Scott}, \citenamefont {Prasannan}, \citenamefont {Dhand},
  \citenamefont {Eigner}, \citenamefont {Quiring}, \citenamefont {Barkhofen},
  \citenamefont {Brecht}, \citenamefont {Plenio},\ and\ \citenamefont
  {Silberhorn}}]{Meyer-Scott2022}%
  \BibitemOpen
  \bibfield  {author} {\bibinfo {author} {\bibfnamefont {E.}~\bibnamefont
  {Meyer-Scott}}, \bibinfo {author} {\bibfnamefont {N.}~\bibnamefont
  {Prasannan}}, \bibinfo {author} {\bibfnamefont {I.}~\bibnamefont {Dhand}},
  \bibinfo {author} {\bibfnamefont {C.}~\bibnamefont {Eigner}}, \bibinfo
  {author} {\bibfnamefont {V.}~\bibnamefont {Quiring}}, \bibinfo {author}
  {\bibfnamefont {S.}~\bibnamefont {Barkhofen}}, \bibinfo {author}
  {\bibfnamefont {B.}~\bibnamefont {Brecht}}, \bibinfo {author} {\bibfnamefont
  {M.~B.}\ \bibnamefont {Plenio}},\ and\ \bibinfo {author} {\bibfnamefont
  {C.}~\bibnamefont {Silberhorn}},\ }\bibfield  {title} {\bibinfo {title}
  {Scalable generation of multiphoton entangled states by active feed-forward
  and multiplexing},\ }\href {https://doi.org/10.1103/PhysRevLett.129.150501}
  {\bibfield  {journal} {\bibinfo  {journal} {Phys. Rev. Lett.}\ }\textbf
  {\bibinfo {volume} {129}},\ \bibinfo {pages} {150501} (\bibinfo {year}
  {2022})}\BibitemShut {NoStop}%
\bibitem [{\citenamefont {Greenberger}\ \emph {et~al.}(1990)\citenamefont
  {Greenberger}, \citenamefont {Horne}, \citenamefont {Shimony},\ and\
  \citenamefont {Zeilinger}}]{Greenberger1990}%
  \BibitemOpen
  \bibfield  {author} {\bibinfo {author} {\bibfnamefont {D.~M.}\ \bibnamefont
  {Greenberger}}, \bibinfo {author} {\bibfnamefont {M.~A.}\ \bibnamefont
  {Horne}}, \bibinfo {author} {\bibfnamefont {A.}~\bibnamefont {Shimony}},\
  and\ \bibinfo {author} {\bibfnamefont {A.}~\bibnamefont {Zeilinger}},\
  }\bibfield  {title} {\bibinfo {title} {{B}ell{'}s theorem without
  inequalities},\ }\href {https://doi.org/10.1119/1.16243} {\bibfield
  {journal} {\bibinfo  {journal} {American Journal of Physics}\ }\textbf
  {\bibinfo {volume} {58}},\ \bibinfo {pages} {1131} (\bibinfo {year}
  {1990})}\BibitemShut {NoStop}%
\bibitem [{\citenamefont {Pan}\ \emph {et~al.}(2000)\citenamefont {Pan},
  \citenamefont {Bouwmeester}, \citenamefont {Daniell}, \citenamefont
  {Weinfurter},\ and\ \citenamefont {Zeilinger}}]{Pan2000}%
  \BibitemOpen
  \bibfield  {author} {\bibinfo {author} {\bibfnamefont {J.-W.}\ \bibnamefont
  {Pan}}, \bibinfo {author} {\bibfnamefont {D.}~\bibnamefont {Bouwmeester}},
  \bibinfo {author} {\bibfnamefont {M.}~\bibnamefont {Daniell}}, \bibinfo
  {author} {\bibfnamefont {H.}~\bibnamefont {Weinfurter}},\ and\ \bibinfo
  {author} {\bibfnamefont {A.}~\bibnamefont {Zeilinger}},\ }\bibfield  {title}
  {\bibinfo {title} {{Experimental test of quantum nonlocality in three-photon
  Greenberger{\textendash}Horne{\textendash}Zeilinger entanglement}},\ }\href
  {https://doi.org/10.1038/35000514} {\bibfield  {journal} {\bibinfo  {journal}
  {Nature}\ }\textbf {\bibinfo {volume} {403}},\ \bibinfo {pages} {515}
  (\bibinfo {year} {2000})}\BibitemShut {NoStop}%
\bibitem [{\citenamefont {Omkar}\ \emph {et~al.}(2022)\citenamefont {Omkar},
  \citenamefont {Lee}, \citenamefont {Teo}, \citenamefont {Lee},\ and\
  \citenamefont {Jeong}}]{Omkar2021}%
  \BibitemOpen
  \bibfield  {author} {\bibinfo {author} {\bibfnamefont {S.}~\bibnamefont
  {Omkar}}, \bibinfo {author} {\bibfnamefont {S.-H.}\ \bibnamefont {Lee}},
  \bibinfo {author} {\bibfnamefont {Y.~S.}\ \bibnamefont {Teo}}, \bibinfo
  {author} {\bibfnamefont {S.-W.}\ \bibnamefont {Lee}},\ and\ \bibinfo {author}
  {\bibfnamefont {H.}~\bibnamefont {Jeong}},\ }\bibfield  {title} {\bibinfo
  {title} {{All-Photonic Architecture for Scalable Quantum Computing with
  Greenberger-Horne-Zeilinger States}},\ }\href
  {https://doi.org/10.1103/PRXQuantum.3.030309} {\bibfield  {journal} {\bibinfo
   {journal} {PRX Quantum}\ }\textbf {\bibinfo {volume} {3}},\ \bibinfo {pages}
  {030309} (\bibinfo {year} {2022})}\BibitemShut {NoStop}%
\bibitem [{\citenamefont {Erhard}\ \emph {et~al.}(2020)\citenamefont {Erhard},
  \citenamefont {Krenn},\ and\ \citenamefont {Zeilinger}}]{Erhard2020}%
  \BibitemOpen
  \bibfield  {author} {\bibinfo {author} {\bibfnamefont {M.}~\bibnamefont
  {Erhard}}, \bibinfo {author} {\bibfnamefont {M.}~\bibnamefont {Krenn}},\ and\
  \bibinfo {author} {\bibfnamefont {A.}~\bibnamefont {Zeilinger}},\ }\bibfield
  {title} {\bibinfo {title} {{Advances in high-dimensional quantum
  entanglement}},\ }\href {https://doi.org/10.1038/s42254-020-0193-5}
  {\bibfield  {journal} {\bibinfo  {journal} {Nat. Rev. Phys.}\ }\textbf
  {\bibinfo {volume} {2}},\ \bibinfo {pages} {365} (\bibinfo {year}
  {2020})}\BibitemShut {NoStop}%
\bibitem [{\citenamefont {Zhang}\ \emph {et~al.}(2020)\citenamefont {Zhang},
  \citenamefont {Liu}, \citenamefont {Zhao}, \citenamefont {Yu},\ and\
  \citenamefont {Yang}}]{Zhang2020}%
  \BibitemOpen
  \bibfield  {author} {\bibinfo {author} {\bibfnamefont {Y.}~\bibnamefont
  {Zhang}}, \bibinfo {author} {\bibfnamefont {T.}~\bibnamefont {Liu}}, \bibinfo
  {author} {\bibfnamefont {J.}~\bibnamefont {Zhao}}, \bibinfo {author}
  {\bibfnamefont {Y.}~\bibnamefont {Yu}},\ and\ \bibinfo {author}
  {\bibfnamefont {C.-P.}\ \bibnamefont {Yang}},\ }\bibfield  {title} {\bibinfo
  {title} {{Generation of hybrid Greenberger-Horne-Zeilinger entangled states
  of particlelike and wavelike optical qubits in circuit QED}},\ }\href
  {https://doi.org/10.1103/PhysRevA.101.062334} {\bibfield  {journal} {\bibinfo
   {journal} {Phys. Rev. A}\ }\textbf {\bibinfo {volume} {101}},\ \bibinfo
  {pages} {062334} (\bibinfo {year} {2020})}\BibitemShut {NoStop}%
\bibitem [{\citenamefont {Xia}\ \emph {et~al.}(2012)\citenamefont {Xia},
  \citenamefont {Chen}, \citenamefont {Song},\ and\ \citenamefont
  {Song}}]{Xia2012}%
  \BibitemOpen
  \bibfield  {author} {\bibinfo {author} {\bibfnamefont {Y.}~\bibnamefont
  {Xia}}, \bibinfo {author} {\bibfnamefont {Q.-Q.}\ \bibnamefont {Chen}},
  \bibinfo {author} {\bibfnamefont {J.}~\bibnamefont {Song}},\ and\ \bibinfo
  {author} {\bibfnamefont {H.-S.}\ \bibnamefont {Song}},\ }\bibfield  {title}
  {\bibinfo {title} {{Efficient hyperentangled
  Greenberger{\textendash}Horne{\textendash}Zeilinger states analysis with
  cross-Kerr nonlinearity}},\ }\href {https://doi.org/10.1364/JOSAB.29.001029}
  {\bibfield  {journal} {\bibinfo  {journal} {J. Opt. Soc. Am. B}\ }\textbf
  {\bibinfo {volume} {29}},\ \bibinfo {pages} {1029} (\bibinfo {year}
  {2012})}\BibitemShut {NoStop}%
\bibitem [{\citenamefont {Maser}\ \emph {et~al.}(2010)\citenamefont {Maser},
  \citenamefont {Wiegner}, \citenamefont {Schilling}, \citenamefont {Thiel},\
  and\ \citenamefont {von Zanthier}}]{Maser2010}%
  \BibitemOpen
  \bibfield  {author} {\bibinfo {author} {\bibfnamefont {A.}~\bibnamefont
  {Maser}}, \bibinfo {author} {\bibfnamefont {R.}~\bibnamefont {Wiegner}},
  \bibinfo {author} {\bibfnamefont {U.}~\bibnamefont {Schilling}}, \bibinfo
  {author} {\bibfnamefont {C.}~\bibnamefont {Thiel}},\ and\ \bibinfo {author}
  {\bibfnamefont {J.}~\bibnamefont {von Zanthier}},\ }\bibfield  {title}
  {\bibinfo {title} {Versatile source of polarization-entangled photons},\
  }\href {https://doi.org/10.1103/PhysRevA.81.053842} {\bibfield  {journal}
  {\bibinfo  {journal} {Phys. Rev. A}\ }\textbf {\bibinfo {volume} {81}},\
  \bibinfo {pages} {053842} (\bibinfo {year} {2010})}\BibitemShut {NoStop}%
\bibitem [{\citenamefont {Bastin}\ \emph {et~al.}(2009)\citenamefont {Bastin},
  \citenamefont {Thiel}, \citenamefont {von Zanthier}, \citenamefont {Lamata},
  \citenamefont {Solano},\ and\ \citenamefont {Agarwal}}]{Bastin2009}%
  \BibitemOpen
  \bibfield  {author} {\bibinfo {author} {\bibfnamefont {T.}~\bibnamefont
  {Bastin}}, \bibinfo {author} {\bibfnamefont {C.}~\bibnamefont {Thiel}},
  \bibinfo {author} {\bibfnamefont {J.}~\bibnamefont {von Zanthier}}, \bibinfo
  {author} {\bibfnamefont {L.}~\bibnamefont {Lamata}}, \bibinfo {author}
  {\bibfnamefont {E.}~\bibnamefont {Solano}},\ and\ \bibinfo {author}
  {\bibfnamefont {G.~S.}\ \bibnamefont {Agarwal}},\ }\bibfield  {title}
  {\bibinfo {title} {Operational determination of multiqubit entanglement
  classes via tuning of local operations},\ }\href
  {https://doi.org/10.1103/PhysRevLett.102.053601} {\bibfield  {journal}
  {\bibinfo  {journal} {Phys. Rev. Lett.}\ }\textbf {\bibinfo {volume} {102}},\
  \bibinfo {pages} {053601} (\bibinfo {year} {2009})}\BibitemShut {NoStop}%
\bibitem [{\citenamefont {Hossein-Nejad}\ \emph {et~al.}(2009)\citenamefont
  {Hossein-Nejad}, \citenamefont {Stock},\ and\ \citenamefont
  {James}}]{Hossein-Nejad2009}%
  \BibitemOpen
  \bibfield  {author} {\bibinfo {author} {\bibfnamefont {H.}~\bibnamefont
  {Hossein-Nejad}}, \bibinfo {author} {\bibfnamefont {R.}~\bibnamefont
  {Stock}},\ and\ \bibinfo {author} {\bibfnamefont {D.~F.~V.}\ \bibnamefont
  {James}},\ }\bibfield  {title} {\bibinfo {title} {{Generation of multiphoton
  entanglement by propagation and detection}},\ }\href
  {https://doi.org/10.1103/PhysRevA.80.022308} {\bibfield  {journal} {\bibinfo
  {journal} {Phys. Rev. A}\ }\textbf {\bibinfo {volume} {80}},\ \bibinfo
  {pages} {022308} (\bibinfo {year} {2009})}\BibitemShut {NoStop}%
\bibitem [{\citenamefont {Pryde}\ and\ \citenamefont
  {White}(2003)}]{Pryde2003}%
  \BibitemOpen
  \bibfield  {author} {\bibinfo {author} {\bibfnamefont {G.~J.}\ \bibnamefont
  {Pryde}}\ and\ \bibinfo {author} {\bibfnamefont {A.~G.}\ \bibnamefont
  {White}},\ }\bibfield  {title} {\bibinfo {title} {Creation of maximally
  entangled photon-number states using optical fiber multiports},\ }\href
  {https://doi.org/10.1103/PhysRevA.68.052315} {\bibfield  {journal} {\bibinfo
  {journal} {Phys. Rev. A}\ }\textbf {\bibinfo {volume} {68}},\ \bibinfo
  {pages} {052315} (\bibinfo {year} {2003})}\BibitemShut {NoStop}%
\bibitem [{\citenamefont {Lim}\ and\ \citenamefont
  {Beige}(2005{\natexlab{a}})}]{Lim2005}%
  \BibitemOpen
  \bibfield  {author} {\bibinfo {author} {\bibfnamefont {Y.~L.}\ \bibnamefont
  {Lim}}\ and\ \bibinfo {author} {\bibfnamefont {A.}~\bibnamefont {Beige}},\
  }\bibfield  {title} {\bibinfo {title} {Multiphoton entanglement through a
  {B}ell-multiport beam splitter},\ }\href
  {https://doi.org/10.1103/PhysRevA.71.062311} {\bibfield  {journal} {\bibinfo
  {journal} {Phys. Rev. A}\ }\textbf {\bibinfo {volume} {71}},\ \bibinfo
  {pages} {062311} (\bibinfo {year} {2005}{\natexlab{a}})}\BibitemShut
  {NoStop}%
\bibitem [{\citenamefont {Kiesel}\ \emph {et~al.}(2010)\citenamefont {Kiesel},
  \citenamefont {Wieczorek}, \citenamefont {Krins}, \citenamefont {Bastin},
  \citenamefont {Weinfurter},\ and\ \citenamefont {Solano}}]{Kiesel2010}%
  \BibitemOpen
  \bibfield  {author} {\bibinfo {author} {\bibfnamefont {N.}~\bibnamefont
  {Kiesel}}, \bibinfo {author} {\bibfnamefont {W.}~\bibnamefont {Wieczorek}},
  \bibinfo {author} {\bibfnamefont {S.}~\bibnamefont {Krins}}, \bibinfo
  {author} {\bibfnamefont {T.}~\bibnamefont {Bastin}}, \bibinfo {author}
  {\bibfnamefont {H.}~\bibnamefont {Weinfurter}},\ and\ \bibinfo {author}
  {\bibfnamefont {E.}~\bibnamefont {Solano}},\ }\bibfield  {title} {\bibinfo
  {title} {{Operational multipartite entanglement classes for symmetric
  photonic qubit states}},\ }\href {https://doi.org/10.1103/PhysRevA.81.032316}
  {\bibfield  {journal} {\bibinfo  {journal} {Phys. Rev. A}\ }\textbf {\bibinfo
  {volume} {81}},\ \bibinfo {pages} {032316} (\bibinfo {year}
  {2010})}\BibitemShut {NoStop}%
\bibitem [{\citenamefont {Kasture}(2018)}]{Kasture2018}%
  \BibitemOpen
  \bibfield  {author} {\bibinfo {author} {\bibfnamefont {S.}~\bibnamefont
  {Kasture}},\ }\bibfield  {title} {\bibinfo {title} {{Scalable approach to
  generation of large symmetric Dicke states}},\ }\href
  {https://doi.org/10.1103/PhysRevA.97.043862} {\bibfield  {journal} {\bibinfo
  {journal} {Phys. Rev. A}\ }\textbf {\bibinfo {volume} {97}},\ \bibinfo
  {pages} {043862} (\bibinfo {year} {2018})}\BibitemShut {NoStop}%
\bibitem [{\citenamefont {Paesani}\ \emph {et~al.}(2021)\citenamefont
  {Paesani}, \citenamefont {Bulmer}, \citenamefont {Jones}, \citenamefont
  {Santagati},\ and\ \citenamefont {Laing}}]{Paesani2021}%
  \BibitemOpen
  \bibfield  {author} {\bibinfo {author} {\bibfnamefont {S.}~\bibnamefont
  {Paesani}}, \bibinfo {author} {\bibfnamefont {J.~F.~F.}\ \bibnamefont
  {Bulmer}}, \bibinfo {author} {\bibfnamefont {A.~E.}\ \bibnamefont {Jones}},
  \bibinfo {author} {\bibfnamefont {R.}~\bibnamefont {Santagati}},\ and\
  \bibinfo {author} {\bibfnamefont {A.}~\bibnamefont {Laing}},\ }\bibfield
  {title} {\bibinfo {title} {Scheme for universal high-dimensional quantum
  computation with linear optics},\ }\href
  {https://doi.org/10.1103/PhysRevLett.126.230504} {\bibfield  {journal}
  {\bibinfo  {journal} {Phys. Rev. Lett.}\ }\textbf {\bibinfo {volume} {126}},\
  \bibinfo {pages} {230504} (\bibinfo {year} {2021})}\BibitemShut {NoStop}%
\bibitem [{\citenamefont {Lee}\ \emph {et~al.}(2022)\citenamefont {Lee},
  \citenamefont {Pramanik}, \citenamefont {Hong}, \citenamefont {Cho},
  \citenamefont {Lim}, \citenamefont {Chin},\ and\ \citenamefont
  {Kim}}]{Lee2022}%
  \BibitemOpen
  \bibfield  {author} {\bibinfo {author} {\bibfnamefont {D.}~\bibnamefont
  {Lee}}, \bibinfo {author} {\bibfnamefont {T.}~\bibnamefont {Pramanik}},
  \bibinfo {author} {\bibfnamefont {S.}~\bibnamefont {Hong}}, \bibinfo {author}
  {\bibfnamefont {Y.-W.}\ \bibnamefont {Cho}}, \bibinfo {author} {\bibfnamefont
  {H.-T.}\ \bibnamefont {Lim}}, \bibinfo {author} {\bibfnamefont
  {S.}~\bibnamefont {Chin}},\ and\ \bibinfo {author} {\bibfnamefont {Y.-S.}\
  \bibnamefont {Kim}},\ }\bibfield  {title} {\bibinfo {title} {Entangling three
  identical particles via spatial overlap},\ }\href
  {https://doi.org/10.1364/OE.460866} {\bibfield  {journal} {\bibinfo
  {journal} {Opt. Express}\ }\textbf {\bibinfo {volume} {30}},\ \bibinfo
  {pages} {30525} (\bibinfo {year} {2022})}\BibitemShut {NoStop}%
\bibitem [{\citenamefont {Bell}\ \emph {et~al.}(2022)\citenamefont {Bell},
  \citenamefont {Bulmer}, \citenamefont {Jones}, \citenamefont {Paesani},
  \citenamefont {McCutcheon},\ and\ \citenamefont {Laing}}]{Bell2022}%
  \BibitemOpen
  \bibfield  {author} {\bibinfo {author} {\bibfnamefont {T.~J.}\ \bibnamefont
  {Bell}}, \bibinfo {author} {\bibfnamefont {J.~F.~F.}\ \bibnamefont {Bulmer}},
  \bibinfo {author} {\bibfnamefont {A.~E.}\ \bibnamefont {Jones}}, \bibinfo
  {author} {\bibfnamefont {S.}~\bibnamefont {Paesani}}, \bibinfo {author}
  {\bibfnamefont {D.~P.~S.}\ \bibnamefont {McCutcheon}},\ and\ \bibinfo
  {author} {\bibfnamefont {A.}~\bibnamefont {Laing}},\ }\bibfield  {title}
  {\bibinfo {title} {{Protocol for generation of high-dimensional entanglement
  from an array of non-interacting photon emitters}},\ }\href
  {https://doi.org/10.1088/1367-2630/ac475d} {\bibfield  {journal} {\bibinfo
  {journal} {New J. Phys.}\ }\textbf {\bibinfo {volume} {24}},\ \bibinfo
  {pages} {013032} (\bibinfo {year} {2022})}\BibitemShut {NoStop}%
\bibitem [{\citenamefont {Ju}\ \emph {et~al.}(2019)\citenamefont {Ju},
  \citenamefont {Yang}, \citenamefont
  {Paunkovi{\ifmmode\acute{c}\else\'{c}\fi}}, \citenamefont {Chu},\ and\
  \citenamefont {Cao}}]{Ju2019}%
  \BibitemOpen
  \bibfield  {author} {\bibinfo {author} {\bibfnamefont {L.}~\bibnamefont
  {Ju}}, \bibinfo {author} {\bibfnamefont {M.}~\bibnamefont {Yang}}, \bibinfo
  {author} {\bibfnamefont {N.}~\bibnamefont
  {Paunkovi{\ifmmode\acute{c}\else\'{c}\fi}}}, \bibinfo {author} {\bibfnamefont
  {W.-J.}\ \bibnamefont {Chu}},\ and\ \bibinfo {author} {\bibfnamefont {Z.-L.}\
  \bibnamefont {Cao}},\ }\bibfield  {title} {\bibinfo {title} {{Creating
  photonic GHZ and W states via quantum walk}},\ }\href
  {https://doi.org/10.1007/s11128-019-2293-7} {\bibfield  {journal} {\bibinfo
  {journal} {Quantum Inf. Process.}\ }\textbf {\bibinfo {volume} {18}},\
  \bibinfo {pages} {176} (\bibinfo {year} {2019})}\BibitemShut {NoStop}%
\bibitem [{\citenamefont {Blasiak}\ and\ \citenamefont
  {Markiewicz}(2019)}]{Blasiak2019}%
  \BibitemOpen
  \bibfield  {author} {\bibinfo {author} {\bibfnamefont {P.}~\bibnamefont
  {Blasiak}}\ and\ \bibinfo {author} {\bibfnamefont {M.}~\bibnamefont
  {Markiewicz}},\ }\bibfield  {title} {\bibinfo {title} {{Entangling three
  qubits without ever touching}},\ }\href
  {https://doi.org/10.1038/s41598-019-55137-3} {\bibfield  {journal} {\bibinfo
  {journal} {Sci. Rep.}\ }\textbf {\bibinfo {volume} {9}},\ \bibinfo {pages}
  {20131} (\bibinfo {year} {2019})}\BibitemShut {NoStop}%
\bibitem [{\citenamefont {Kim}\ \emph {et~al.}(2020)\citenamefont {Kim},
  \citenamefont {Cho}, \citenamefont {Lim},\ and\ \citenamefont
  {Han}}]{Kim2020}%
  \BibitemOpen
  \bibfield  {author} {\bibinfo {author} {\bibfnamefont {Y.-S.}\ \bibnamefont
  {Kim}}, \bibinfo {author} {\bibfnamefont {Y.-W.}\ \bibnamefont {Cho}},
  \bibinfo {author} {\bibfnamefont {H.-T.}\ \bibnamefont {Lim}},\ and\ \bibinfo
  {author} {\bibfnamefont {S.-W.}\ \bibnamefont {Han}},\ }\bibfield  {title}
  {\bibinfo {title} {{Efficient linear optical generation of a multipartite $W$
  state via a quantum eraser}},\ }\href
  {https://doi.org/10.1103/PhysRevA.101.022337} {\bibfield  {journal} {\bibinfo
   {journal} {Phys. Rev. A}\ }\textbf {\bibinfo {volume} {101}},\ \bibinfo
  {pages} {022337} (\bibinfo {year} {2020})}\BibitemShut {NoStop}%
\bibitem [{\citenamefont {Blasiak}\ \emph {et~al.}(2021)\citenamefont
  {Blasiak}, \citenamefont {Borsuk}, \citenamefont {Markiewicz},\ and\
  \citenamefont {Kim}}]{Blasiak2021}%
  \BibitemOpen
  \bibfield  {author} {\bibinfo {author} {\bibfnamefont {P.}~\bibnamefont
  {Blasiak}}, \bibinfo {author} {\bibfnamefont {E.}~\bibnamefont {Borsuk}},
  \bibinfo {author} {\bibfnamefont {M.}~\bibnamefont {Markiewicz}},\ and\
  \bibinfo {author} {\bibfnamefont {Y.-S.}\ \bibnamefont {Kim}},\ }\bibfield
  {title} {\bibinfo {title} {Efficient linear-optical generation of a
  multipartite ${W}$ state},\ }\href
  {https://doi.org/10.1103/PhysRevA.104.023701} {\bibfield  {journal} {\bibinfo
   {journal} {Phys. Rev. A}\ }\textbf {\bibinfo {volume} {104}},\ \bibinfo
  {pages} {023701} (\bibinfo {year} {2021})}\BibitemShut {NoStop}%
\bibitem [{\citenamefont {Kumar}\ \emph {et~al.}(2022)\citenamefont {Kumar},
  \citenamefont {Bhatti}, \citenamefont {Jones},\ and\ \citenamefont
  {Barz}}]{Kumar2022}%
  \BibitemOpen
  \bibfield  {author} {\bibinfo {author} {\bibfnamefont {S.}~\bibnamefont
  {Kumar}}, \bibinfo {author} {\bibfnamefont {D.}~\bibnamefont {Bhatti}},
  \bibinfo {author} {\bibfnamefont {A.~E.}\ \bibnamefont {Jones}},\ and\
  \bibinfo {author} {\bibfnamefont {S.}~\bibnamefont {Barz}},\ }\bibfield
  {title} {\bibinfo {title} {Entanglement generation using multiport
  beam splitters},\ }\href@noop {} {\bibfield  {journal} {\bibinfo  {journal}
  {in preparation}\ } (\bibinfo {year} {2022})}\BibitemShut {NoStop}%
\bibitem [{\citenamefont {Menssen}\ \emph {et~al.}(2017)\citenamefont
  {Menssen}, \citenamefont {Jones}, \citenamefont {Metcalf}, \citenamefont
  {Tichy}, \citenamefont {Barz}, \citenamefont {Kolthammer},\ and\
  \citenamefont {Walmsley}}]{Menssen2017}%
  \BibitemOpen
  \bibfield  {author} {\bibinfo {author} {\bibfnamefont {A.~J.}\ \bibnamefont
  {Menssen}}, \bibinfo {author} {\bibfnamefont {A.~E.}\ \bibnamefont {Jones}},
  \bibinfo {author} {\bibfnamefont {B.~J.}\ \bibnamefont {Metcalf}}, \bibinfo
  {author} {\bibfnamefont {M.~C.}\ \bibnamefont {Tichy}}, \bibinfo {author}
  {\bibfnamefont {S.}~\bibnamefont {Barz}}, \bibinfo {author} {\bibfnamefont
  {W.~S.}\ \bibnamefont {Kolthammer}},\ and\ \bibinfo {author} {\bibfnamefont
  {I.~A.}\ \bibnamefont {Walmsley}},\ }\bibfield  {title} {\bibinfo {title}
  {Distinguishability and many-particle interference},\ }\href
  {https://doi.org/10.1103/PhysRevLett.118.153603} {\bibfield  {journal}
  {\bibinfo  {journal} {Phys. Rev. Lett.}\ }\textbf {\bibinfo {volume} {118}},\
  \bibinfo {pages} {153603} (\bibinfo {year} {2017})}\BibitemShut {NoStop}%
\bibitem [{\citenamefont {Dittel}\ \emph {et~al.}(2018)\citenamefont {Dittel},
  \citenamefont {Dufour}, \citenamefont {Walschaers}, \citenamefont {Weihs},
  \citenamefont {Buchleitner},\ and\ \citenamefont {Keil}}]{Dittel2018}%
  \BibitemOpen
  \bibfield  {author} {\bibinfo {author} {\bibfnamefont {C.}~\bibnamefont
  {Dittel}}, \bibinfo {author} {\bibfnamefont {G.}~\bibnamefont {Dufour}},
  \bibinfo {author} {\bibfnamefont {M.}~\bibnamefont {Walschaers}}, \bibinfo
  {author} {\bibfnamefont {G.}~\bibnamefont {Weihs}}, \bibinfo {author}
  {\bibfnamefont {A.}~\bibnamefont {Buchleitner}},\ and\ \bibinfo {author}
  {\bibfnamefont {R.}~\bibnamefont {Keil}},\ }\bibfield  {title} {\bibinfo
  {title} {Totally destructive interference for permutation-symmetric
  many-particle states},\ }\href {https://doi.org/10.1103/PhysRevA.97.062116}
  {\bibfield  {journal} {\bibinfo  {journal} {Phys. Rev. A}\ }\textbf {\bibinfo
  {volume} {97}},\ \bibinfo {pages} {062116} (\bibinfo {year}
  {2018})}\BibitemShut {NoStop}%
\bibitem [{\citenamefont {Jones}\ \emph {et~al.}(2020)\citenamefont {Jones},
  \citenamefont {Menssen}, \citenamefont {Chrzanowski}, \citenamefont
  {Wolterink}, \citenamefont {Shchesnovich},\ and\ \citenamefont
  {Walmsley}}]{Jones2020}%
  \BibitemOpen
  \bibfield  {author} {\bibinfo {author} {\bibfnamefont {A.~E.}\ \bibnamefont
  {Jones}}, \bibinfo {author} {\bibfnamefont {A.~J.}\ \bibnamefont {Menssen}},
  \bibinfo {author} {\bibfnamefont {H.~M.}\ \bibnamefont {Chrzanowski}},
  \bibinfo {author} {\bibfnamefont {T.~A.~W.}\ \bibnamefont {Wolterink}},
  \bibinfo {author} {\bibfnamefont {V.~S.}\ \bibnamefont {Shchesnovich}},\ and\
  \bibinfo {author} {\bibfnamefont {I.~A.}\ \bibnamefont {Walmsley}},\
  }\bibfield  {title} {\bibinfo {title} {{Multiparticle Interference of
  Pairwise Distinguishable Photons}},\ }\href
  {https://doi.org/10.1103/PhysRevLett.125.123603} {\bibfield  {journal}
  {\bibinfo  {journal} {Phys. Rev. Lett.}\ }\textbf {\bibinfo {volume} {125}},\
  \bibinfo {pages} {123603} (\bibinfo {year} {2020})}\BibitemShut {NoStop}%
\bibitem [{\citenamefont {Minke}\ \emph {et~al.}(2021)\citenamefont {Minke},
  \citenamefont {Buchleitner},\ and\ \citenamefont {Dittel}}]{Minke2021}%
  \BibitemOpen
  \bibfield  {author} {\bibinfo {author} {\bibfnamefont {A.~M.}\ \bibnamefont
  {Minke}}, \bibinfo {author} {\bibfnamefont {A.}~\bibnamefont {Buchleitner}},\
  and\ \bibinfo {author} {\bibfnamefont {C.}~\bibnamefont {Dittel}},\
  }\bibfield  {title} {\bibinfo {title} {{Characterizing four-body
  indistinguishability via symmetries}},\ }\href
  {https://doi.org/10.1088/1367-2630/ac0fb1} {\bibfield  {journal} {\bibinfo
  {journal} {New J. Phys.}\ }\textbf {\bibinfo {volume} {23}},\ \bibinfo
  {pages} {073028} (\bibinfo {year} {2021})}\BibitemShut {NoStop}%
\bibitem [{\citenamefont {Shih}\ and\ \citenamefont {Alley}(1988)}]{Shih1988}%
  \BibitemOpen
  \bibfield  {author} {\bibinfo {author} {\bibfnamefont {Y.~H.}\ \bibnamefont
  {Shih}}\ and\ \bibinfo {author} {\bibfnamefont {C.~O.}\ \bibnamefont
  {Alley}},\ }\bibfield  {title} {\bibinfo {title} {{New Type of
  Einstein-Podolsky-Rosen-Bohm Experiment Using Pairs of Light Quanta Produced
  by Optical Parametric Down Conversion}},\ }\href
  {https://doi.org/10.1103/PhysRevLett.61.2921} {\bibfield  {journal} {\bibinfo
   {journal} {Phys. Rev. Lett.}\ }\textbf {\bibinfo {volume} {61}},\ \bibinfo
  {pages} {2921} (\bibinfo {year} {1988})}\BibitemShut {NoStop}%
\bibitem [{\citenamefont {Lim}\ and\ \citenamefont
  {Beige}(2005{\natexlab{b}})}]{Lim2005IOP}%
  \BibitemOpen
  \bibfield  {author} {\bibinfo {author} {\bibfnamefont {Y.~L.}\ \bibnamefont
  {Lim}}\ and\ \bibinfo {author} {\bibfnamefont {A.}~\bibnamefont {Beige}},\
  }\bibfield  {title} {\bibinfo {title} {Generalized
  {H}ong{\textendash}{O}u{\textendash}{M}andel experiments with bosons and
  fermions},\ }\href {https://doi.org/10.1088/1367-2630/7/1/155} {\bibfield
  {journal} {\bibinfo  {journal} {New J. Phys.}\ }\textbf {\bibinfo
  {volume} {7}},\ \bibinfo {pages} {155} (\bibinfo {year}
  {2005}{\natexlab{b}})}\BibitemShut {NoStop}%
\bibitem [{\citenamefont {Tichy}\ \emph {et~al.}(2010)\citenamefont {Tichy},
  \citenamefont {Tiersch}, \citenamefont {de~Melo}, \citenamefont {Mintert},\
  and\ \citenamefont {Buchleitner}}]{Tichy2010}%
  \BibitemOpen
  \bibfield  {author} {\bibinfo {author} {\bibfnamefont {M.~C.}\ \bibnamefont
  {Tichy}}, \bibinfo {author} {\bibfnamefont {M.}~\bibnamefont {Tiersch}},
  \bibinfo {author} {\bibfnamefont {F.}~\bibnamefont {de~Melo}}, \bibinfo
  {author} {\bibfnamefont {F.}~\bibnamefont {Mintert}},\ and\ \bibinfo {author}
  {\bibfnamefont {A.}~\bibnamefont {Buchleitner}},\ }\bibfield  {title}
  {\bibinfo {title} {{Zero-Transmission Law for Multiport Beam Splitters}},\
  }\href {https://doi.org/10.1103/PhysRevLett.104.220405} {\bibfield  {journal}
  {\bibinfo  {journal} {Phys. Rev. Lett.}\ }\textbf {\bibinfo {volume} {104}},\
  \bibinfo {pages} {220405} (\bibinfo {year} {2010})}\BibitemShut {NoStop}%
\bibitem [{\citenamefont {Classen}\ \emph {et~al.}(2016)\citenamefont
  {Classen}, \citenamefont {Waldmann}, \citenamefont {Giebel}, \citenamefont
  {Schneider}, \citenamefont {Bhatti}, \citenamefont {Mehringer},\ and\
  \citenamefont {von Zanthier}}]{Classen2016}%
  \BibitemOpen
  \bibfield  {author} {\bibinfo {author} {\bibfnamefont {A.}~\bibnamefont
  {Classen}}, \bibinfo {author} {\bibfnamefont {F.}~\bibnamefont {Waldmann}},
  \bibinfo {author} {\bibfnamefont {S.}~\bibnamefont {Giebel}}, \bibinfo
  {author} {\bibfnamefont {R.}~\bibnamefont {Schneider}}, \bibinfo {author}
  {\bibfnamefont {D.}~\bibnamefont {Bhatti}}, \bibinfo {author} {\bibfnamefont
  {T.}~\bibnamefont {Mehringer}},\ and\ \bibinfo {author} {\bibfnamefont
  {J.}~\bibnamefont {von Zanthier}},\ }\bibfield  {title} {\bibinfo {title}
  {Superresolving imaging of arbitrary one-dimensional arrays of thermal light
  sources using multiphoton interference},\ }\href
  {https://doi.org/10.1103/PhysRevLett.117.253601} {\bibfield  {journal}
  {\bibinfo  {journal} {Phys. Rev. Lett.}\ }\textbf {\bibinfo {volume} {117}},\
  \bibinfo {pages} {253601} (\bibinfo {year} {2016})}\BibitemShut {NoStop}%
\bibitem [{\citenamefont {Bhatti}\ \emph {et~al.}(2018)\citenamefont {Bhatti},
  \citenamefont {Classen}, \citenamefont {Oppel}, \citenamefont {Schneider},\
  and\ \citenamefont {von Zanthier}}]{Bhatti2018}%
  \BibitemOpen
  \bibfield  {author} {\bibinfo {author} {\bibfnamefont {D.}~\bibnamefont
  {Bhatti}}, \bibinfo {author} {\bibfnamefont {A.}~\bibnamefont {Classen}},
  \bibinfo {author} {\bibfnamefont {S.}~\bibnamefont {Oppel}}, \bibinfo
  {author} {\bibfnamefont {R.}~\bibnamefont {Schneider}},\ and\ \bibinfo
  {author} {\bibfnamefont {J.}~\bibnamefont {von Zanthier}},\ }\bibfield
  {title} {\bibinfo {title} {{Generation of N00N-like interferences with two
  thermal light sources}},\ }\href {https://doi.org/10.1140/epjd/e2018-90371-8}
  {\bibfield  {journal} {\bibinfo  {journal} {Eur. Phys. J. D}\ }\textbf
  {\bibinfo {volume} {72}},\ \bibinfo {pages} {191} (\bibinfo {year}
  {2018})}\BibitemShut {NoStop}%
\bibitem [{\citenamefont {Aaronson}\ and\ \citenamefont
  {Arkhipov}(2011)}]{Aaronson2011}%
  \BibitemOpen
  \bibfield  {author} {\bibinfo {author} {\bibfnamefont {S.}~\bibnamefont
  {Aaronson}}\ and\ \bibinfo {author} {\bibfnamefont {A.}~\bibnamefont
  {Arkhipov}},\ }\bibfield  {title} {\bibinfo {title} {{The computational
  complexity of linear optics}},\ }in\ \href
  {https://doi.org/10.1145/1993636.1993682} {\emph {\bibinfo {booktitle} {{STOC
  '11: Proceedings of the forty-third annual ACM symposium on Theory of
  computing}}}}\ (\bibinfo  {publisher} {ACM},\ \bibinfo {address} {New York},\
  \bibinfo {year} {2011})\ pp.\ \bibinfo {pages} {333--342}\BibitemShut
  {NoStop}%
\bibitem [{\citenamefont {Sagi}(2003)}]{Sagi2003}%
  \BibitemOpen
  \bibfield  {author} {\bibinfo {author} {\bibfnamefont {Y.}~\bibnamefont
  {Sagi}},\ }\bibfield  {title} {\bibinfo {title} {{Scheme for generating
  Greenberger-Horne-Zeilinger-type states of $n$ photons}},\ }\href
  {https://doi.org/10.1103/PhysRevA.68.042320} {\bibfield  {journal} {\bibinfo
  {journal} {Phys. Rev. A}\ }\textbf {\bibinfo {volume} {68}},\ \bibinfo
  {pages} {042320} (\bibinfo {year} {2003})}\BibitemShut {NoStop}%
\bibitem [{\citenamefont {Gimeno-Segovia}\ \emph {et~al.}(2015)\citenamefont
  {Gimeno-Segovia}, \citenamefont {Shadbolt}, \citenamefont {Browne},\ and\
  \citenamefont {Rudolph}}]{Gimeno-Segovia2015}%
  \BibitemOpen
  \bibfield  {author} {\bibinfo {author} {\bibfnamefont {M.}~\bibnamefont
  {Gimeno-Segovia}}, \bibinfo {author} {\bibfnamefont {P.}~\bibnamefont
  {Shadbolt}}, \bibinfo {author} {\bibfnamefont {D.~E.}\ \bibnamefont
  {Browne}},\ and\ \bibinfo {author} {\bibfnamefont {T.}~\bibnamefont
  {Rudolph}},\ }\bibfield  {title} {\bibinfo {title} {{From Three-Photon
  Greenberger-Horne-Zeilinger States to Ballistic Universal Quantum
  Computation}},\ }\href {https://doi.org/10.1103/PhysRevLett.115.020502}
  {\bibfield  {journal} {\bibinfo  {journal} {Phys. Rev. Lett.}\ }\textbf
  {\bibinfo {volume} {115}},\ \bibinfo {pages} {020502} (\bibinfo {year}
  {2015})}\BibitemShut {NoStop}%
\bibitem [{\citenamefont {Jones}\ \emph {et~al.}(2022)\citenamefont {Jones},
  \citenamefont {Kumar}, \citenamefont {D'Aurelio}, \citenamefont {Bayerbach},
  \citenamefont {Menssen},\ and\ \citenamefont {Barz}}]{Jones2022}%
  \BibitemOpen
  \bibfield  {author} {\bibinfo {author} {\bibfnamefont {A.~E.}\ \bibnamefont
  {Jones}}, \bibinfo {author} {\bibfnamefont {S.}~\bibnamefont {Kumar}},
  \bibinfo {author} {\bibfnamefont {S.}~\bibnamefont {D'Aurelio}}, \bibinfo
  {author} {\bibfnamefont {M.}~\bibnamefont {Bayerbach}}, \bibinfo {author}
  {\bibfnamefont {A.~J.}\ \bibnamefont {Menssen}},\ and\ \bibinfo {author}
  {\bibfnamefont {S.}~\bibnamefont {Barz}},\ }\bibfield  {title} {\bibinfo
  {title} {{Distinguishability and mixedness in quantum interference}},\ }\href
  {https://doi.org/10.48550/arXiv.2201.04655} {\bibfield  {journal} {\bibinfo
  {journal} {arXiv:2201.04655}\ } (\bibinfo {year} {2022})}\BibitemShut
  {NoStop}%
\bibitem [{\citenamefont {Hofmann}(2004)}]{Hofmann2004}%
  \BibitemOpen
  \bibfield  {author} {\bibinfo {author} {\bibfnamefont {H.~F.}\ \bibnamefont
  {Hofmann}},\ }\bibfield  {title} {\bibinfo {title} {Generation of highly
  nonclassical $n$-photon polarization states by superbunching at a photon
  bottleneck},\ }\href {https://doi.org/10.1103/PhysRevA.70.023812} {\bibfield
  {journal} {\bibinfo  {journal} {Phys. Rev. A}\ }\textbf {\bibinfo {volume}
  {70}},\ \bibinfo {pages} {023812} (\bibinfo {year} {2004})}\BibitemShut
  {NoStop}%
\bibitem [{\citenamefont {Seron}\ \emph {et~al.}(2022)\citenamefont {Seron},
  \citenamefont {Novo},\ and\ \citenamefont {Cerf}}]{Seron2022}%
  \BibitemOpen
  \bibfield  {author} {\bibinfo {author} {\bibfnamefont {B.}~\bibnamefont
  {Seron}}, \bibinfo {author} {\bibfnamefont {L.}~\bibnamefont {Novo}},\ and\
  \bibinfo {author} {\bibfnamefont {N.~J.}\ \bibnamefont {Cerf}},\ }\bibfield
  {title} {\bibinfo {title} {{Boson bunching is not maximized by
  indistinguishable particles}},\ }\href
  {https://doi.org/10.48550/arXiv.2203.01306} {\bibfield  {journal} {\bibinfo
  {journal} {arXiv:2203.01306}\ } (\bibinfo {year} {2022})}\BibitemShut
  {NoStop}%
\end{thebibliography}
\end{document}